\documentclass[12pt]{article}
\usepackage{graphicx}
\usepackage{natbib} 
\usepackage{url} 
\usepackage{amsmath, amssymb, amsfonts, amsthm, float}  
\usepackage{enumerate, color, framed, float, multirow,subcaption}

\newcommand{\blind}{0}

\allowdisplaybreaks

\newcommand{\BR}{{\mathbb R}}

\definecolor{Gray}{gray}{0.9}

\theoremstyle{remark}

\begin{document}
\def\spacingset#1{\renewcommand{\baselinestretch}%
{#1}\small\normalsize} \spacingset{1}

\if0\blind
{
  \title{\bf Bayesian equation selection on sparse data for discovery of stochastic dynamical systems}

  \author{Kushagra Gupta\hspace{.2cm}\\
    Department of Mathematics and Statistics\\ Indian Institute of Technology Kanpur\\
    and \\
    Dootika Vats \\
    Department of Mathematics and Statistics\\ Indian Institute of Technology Kanpur\\
    and \\
    Snigdhansu Chatterjee \\ 
    School of Statistics\\ University of Minnesota}
  \maketitle
} \fi

\if1\blind
{
  \bigskip
  \bigskip
  \bigskip
  \begin{center}
    {\LARGE\bf Title}
\end{center}
  \medskip
} \fi

\bigskip
\begin{abstract}


Often the underlying system of differential equations driving a stochastic dynamical system is assumed to be known, with inference conditioned on this assumption. 
We present a Bayesian framework for discovering this  system of differential equations under assumptions that align with real-life scenarios, including the availability of relatively sparse data. Further, we discuss computational strategies that are critical in teasing out the important details about the dynamical system  and algorithmic innovations to solve for acute parameter interdependence in the absence of rich data. This gives a complete Bayesian pathway for model identification via a variable selection paradigm and  parameter estimation of the corresponding model using only the observed data. We present detailed computations and analysis of the Lorenz-96, Lorenz-63, and the Orstein-Uhlenbeck system using the Bayesian framework we propose. 
\end{abstract}

\noindent%
\vfill

\newpage
\spacingset{2} 

\section{Introduction} 
\label{sec:introduction}


In the present era of data-driven scientific discoveries,  consequent industrial and commercial applications, and new scientific hypothesis generation, it is useful to understand how experimental and observed data may inform  scientists about underlying physical systems. In this paper, we consider data arising from 
\textbf{\textit{stochastic dynamical systems}},  and present a detailed and thorough Bayesian route-map of the process of scientific discoveries from noisy data, along with algorithmic innovations that greatly facilitate the computational pathways leading to such discoveries. Along the way, we provide honest discussions on the pitfalls, limitations, and computational challenges of trying to discover ``science from data''. 

More precisely, we consider a stochastic dynamic system given by 
\begin{equation}
	d \mathbf{X}_{t} = f(t, \mathbf{X}_{t} ; \theta) + \Sigma^{1/2} \, d \mathbf{W}_{t} \, ,
	\label{eq:SDE}
\end{equation}
where $f (\cdot) \in \BR^{p}$ is the drift function with parameter $\theta$, $\Sigma$ is the $p \times p$ diffusion covariance matrix, and $\mathbf{W}_{t}$ is a standard $p$-dimensional Wiener process. Such dynamical systems are employed in a vast number of natural and social science disciplines, including atmosphere, ocean and other climate and weather systems modeling, systems and control theory, flows and turbulence, molecular dynamics, cancer and other translational biological systems, econometrics and finance, modeling of behavioral and cognitive processes, and so on. For clarity of presentation and precise formulation, in this paper we discuss the Lorenz-96 \citep{lorenz:1996}, the Lorenz-63 \citep{lorenz:1963} and the Ornstein-Uhlenbeck (OU) \citep{uhlen:orste:1930} processes in detail. Further, we assume 
that the continuous process $\{ \mathbf{X}_{t}: t \in \BR \}$ is latent and unobserved, and the observed noisy data is 
\begin{equation}
\label{eq:data_Y}
	\mathbf{Y}_{i} = \mathbf{X}_{t_{i}} +  \epsilon_{i}, \,  \quad i =1, \ldots, K;
\end{equation}
with $\{ t_{i}\}_{i=1}^{K}$ being the time-stamps of the associated unobserved latent process $\{ \mathbf{X}_{t} \}$. The  $\{ \epsilon_{i}\}_{i=1}^{K}$ are additive noise in the observed data, which for clarity, we consider to be independent and identically distributed as $p$-dimensional Gaussian random variables with mean zero and known variance 
$R$. That is, $\epsilon_{i} \overset{\text{iid}}{\sim} \ N_{p} (0, R)$.  The $p \times p$ covariance matrix parameter, $R$, is assumed to be known for identifiability. In keeping with real application scenarios, we assume that $K$ is not too large, hence we essentially have \textit{sparse}
data for the scientific modeling process. 

In related literature in the physical sciences, the functional form of $f (\cdot)$ is typically assumed known up to a finite number of parameters, and the computational steps and heuristics employed often imply that $\{ \mathbf{X}_{t} \}$ is effectively known and fixed, thus resulting in considerably simple computations and underestimation of the uncertainties. Here, we propose a study where neither the drift function nor the latent process  is known. 

\begin{figure}[htbp]
  \centering
  \includegraphics[width = 0.45\textwidth]{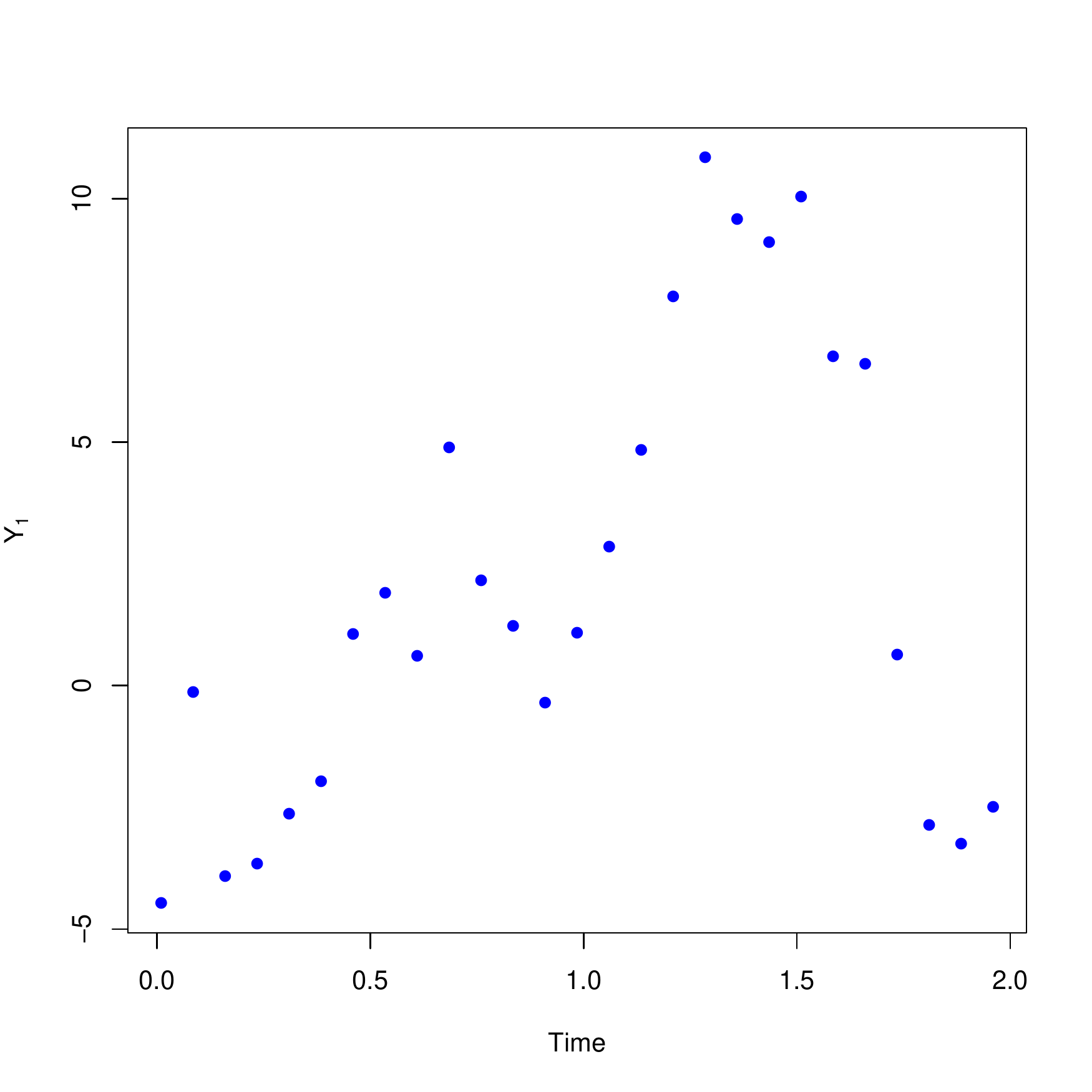}
  \includegraphics[width = 0.45\textwidth]{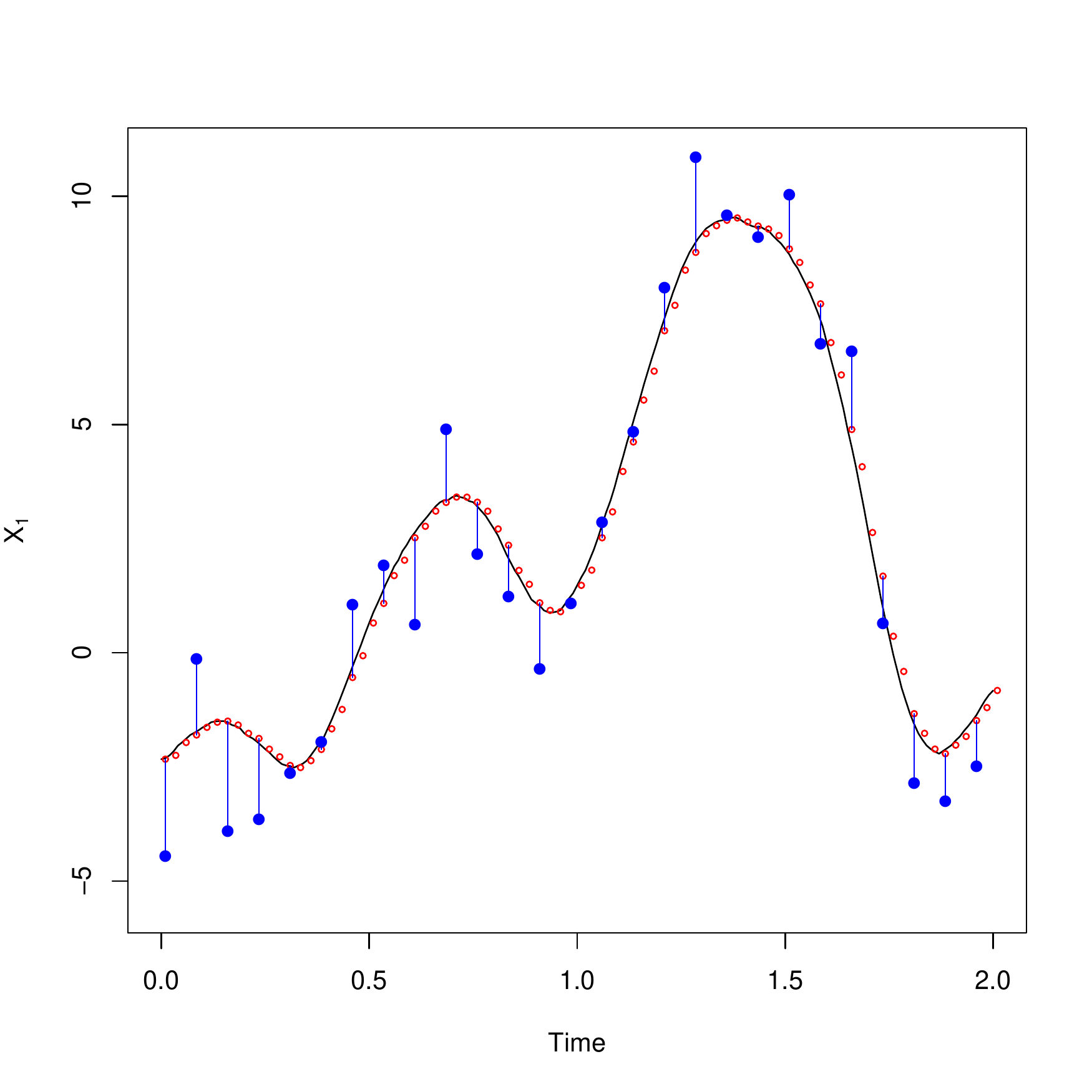}
  \caption{A pictorial representation of the first component of a 4-dimensional Lorenz 96 system. The left panel is the observed data in a time interval $[0,2]$. The right panel contains the same data, but also the latent underlying  continuous process (black curve), and it's discretized version (red dots) over a uniform grid of points.}
  \label{fig:lorenz96data}
\end{figure}

Eliciting the drift function $f (\cdot)$, estimating the unknown parameters: $\theta$ and  $\Sigma$, and predicting the unobserved continuous process 
$\{ \mathbf{X}_{t} \}$ are some of the challenges that we address in this paper. 
Our contributions are threefold: (i) we present a  Bayesian variable selection technique for identification and elicitation of the underlying dynamic systems (ii) we discuss the sources of unreliability and instability  in inferring dynamical systems from observed data, and (iii) we present computational strategies that are critical in teasing out the important details about the dynamical system from sparse 
observations. The following example serves as an illustration:


\textsc{Example:}
Suppose an atmospheric scientist observes a $p = 4$ dimensional time series data, the first component  of which is depicted in the left panel of Figure~\ref{fig:lorenz96data}. 
It is  surmised that the latent continuous atmospheric diffusion process driving this data has a temporal evolution system that can be reasonably approximated by a  low-order polynomial of the state variables. It is possible that the current scientific knowledge informs or conjectures about some terms of this polynomial, or some relations between the polynomial coefficients, and can serve as a starting point for the construction of a \textit{digital twin} system for the actual stochastic dynamical system.

The Bayesian modeling framework that we propose addresses the above scientific questions 
using a scheme illustrated on the right panel of Figure~\ref{fig:lorenz96data}. The unobserved latent continuous process is denoted by the black curve, the red hollow dots denote a discretized version of the same, and the blue solid dots are the observed data. The goal of this paper is essentially to tease out the equation of the black curve from the blue dots. 
%
\qed

There are multiple challenges in this process. 
Notice that the complexity of the problem and sparsity of data are related to the assumed discretization structure (red hollow dots in the right panel of Figure~\ref{fig:lorenz96data}) of
the continuous latent process $\{ \mathbf{X}_{t} \}$ (black curve in the right panel of Figure~\ref{fig:lorenz96data}). If we adopt too fine a grid, the discrete representation better captures the properties of the actual latent stochastic process, but leads to
a high-dimensional discrete latent variable and considerable computational difficulty, as will be apparent later. A coarse discretization results in a poor representation of the underlying actual stochastic process, and unreliable inference and prediction. 

Once a discretization framework is fixed, the first step is to construct a Bayesian variable or model  selection framework that allows scientists to choose an appropriate dynamical system from a given dictionary of physically meaningful and interpretable systems. This generally results in selecting an appropriate functional form of the drift function, $f (\cdot)$, from a parametric family of functions. For this, we propose a spike-and-slab prior driven model selection framework. To the best of our knowledge, 
using Bayesian model selection for eliciting the structure of a dynamical system, or
``\textit{stochastic ordinary/partial differential equation selection}'' has not been attempted before.


Once the functional form  of $f (\cdot)$  is decided,  the estimation of $\theta$ and  $\Sigma$, and prediction of $\mathbf{X}$ are  the next challenges to address. 
The posterior distribution of $(\theta, \Sigma, \mathbf{X})$ is particularly complex, because  $\mathbf{X}$ is high-dimensional, and because there is high correlation within the $\mathbf{X}$s, as well as large dependency between the $\mathbf{X}$s and $\Sigma$. 
Also, as noted by \cite{vrettas:cornford:2011}, the high dependency between the $\mathbf{X}$s and $\Sigma$ yields slow mixing chains in standard Markov chain Monte Carlo (MCMC often hereafter) algorithms. However, we notice that $\Sigma$ can be integrated out of the joint posterior, and this observation leads us to implement a \textbf{\textit{linchpin variable MCMC sampler}} as described by \cite{archila:2016}. This marginalization yields significant improvements in the mixing of the Markov chain, implying improved quality of estimation.  



Current research on studying data from stochastic dynamical systems may be found both in Statistics as well as Physics literature, however, using data in a principled way to \textit{discover} the underlying stochastic differential equation structure seems new. 
The existing literature in both Statistics and Physics overwhelmingly use Bayesian methods, but there are important differences. The statistical literature includes theoretical and methodological advances, and may be sampled 
from \cite{bhaumik2014bayesian,brunel2008parameter, chkrebtii2016bayesian,hennig2015probabilistic, lie2018random, matsuda2019estimation,  ramsay2007parameter,  wang2020role, zhang2017bayesian} and the numerous papers related to these. While 
there are important scientific problems studied in these papers, we believe our context 
involving sparse data and unknown drift function is new, as are the computational advancements that we propose.

A sample of the related literature from the physical sciences may be found in  \cite{ala2015gaussian,apte2007sampling, batz2018approximate,  ching2006bayesian, perez2018probabilistic,  vrettas:cornford:2011, 
vrettas2010new}. However, we noticed a few 
possibly significant gaps in statistical treatment in this line of work.
Often, to facilitate computations or illustrate the main scientific ideas, \textit{ad hoc} approximations and assumptions are made, and heuristics are employed. 
For example, the drift function 
$f (\cdot)$ may be assumed to be linear or approximated by a linear function. Additionally, the numeric computations may imply that $\{ \mathbf{X}_{t} \}$ is effectively fixed and known (and thus neither random nor latent) for the actual Bayesian computation. This is a critical difference, since the uncertainty due to 
$\{ \mathbf{X}_{t} \}$ is not reflected in the estimation or inference steps.

Recently, in order to tackle the high-dimensional nature of the estimation problem, there has been significant work in the use of variational inference algorithms to approximate the posterior distribution. This includes the works of \citet{archambeau2007variational,bau:gor:2017,yu:Li:Xu:2018} and many others. As we will demonstrate, in many chaotic systems, a small deviation away from the underlying true parameters can have dramatic effects on  inference for such models. Thus, approximations to the true posterior distribution with no quantification of the approximation error (as is the case with variational Bayes algorithms), can lead to dramatically unreliable inference.

In a different direction, \cite{eraker:2001} propose a component-wise MCMC algorithm for discretized univariate diffusions. However, the high dependence between the $\mathbf{X}$s yields large auto-correlation and high cross-correlation in the Markov chain.  
When the functional form $f (\cdot)$ is known but involves many parameters with adequate available data, and a Euler-Maruyama approximation is not required for the statistical model, the approach due to \cite{zhang2017bayesian} may be used.

In Section~\ref{sec:bayesian_model} below, we present in detail the spike-and-slab model that we  adopt in this paper. Our proposed Bayesian algorithm and the computational aspects are described in Section~\ref{sec:computation}.
In Section~\ref{sec:examples}, we analyze the performance of our proposed dynamic stochastic model selection paradigm and subsequent inference for data obtained by three systems: a Lorenz 96, an Ornstein-Uhlenbeck process, and a particularly chaotic Lorenz 63 model. In all three examples, our model is successfully able to identify the underlying dynamical process. We also draw attention to the performance of MCMC when reasonable starting values of the latent paths is unavailable and caution practitioners against short MCMC runs often witnessed in the literature. Our conclusions are summarized in 
Section~\ref{sec:future_work}. Additional technical details are included in the supplementary materials.



\section{Spike-and-slab model for system identification}
\label{sec:bayesian_model}

Recall that $t_1$ and  $t_K$ are the time-stamps for the first and last observations. We first use an Euler-Maruyama discretization step on $\{ \mathbf{X}_t \}$, 
to obtain the latent vector $\mathbf{X} = (\mathbf{X}_0, \dots, \mathbf{X}_{N})$, where 
we select a grid-size $\delta t$ and define $N = (t_K - t_1)/\delta t$. The discrete latent vector ($\mathbf{X}$) and the continuous latent process ($\{ \mathbf{X}_t \}$) have similar but distinct notations, to signify their close correspondence. Based on \eqref{eq:SDE}, this discretization yields the Markovian structure
 \begin{equation*}
	\mathbf{X}_{k+1} = \mathbf{X}_{k} + f(t, \, \mathbf{X}_{k}; \theta)\delta t + \sqrt{\Sigma \delta t}\,\xi_{k} \, ,
 \end{equation*}
where $\xi_{k} \sim N_p(0,I_p)$  for $k = 0, \dots, N-1$. Thus, the joint density of $\mathbf{X}$ satisfies
\[
 \pi(\mathbf{X}) = \pi(\mathbf{X}_{0}) \, \prod_{i=1}^{N} \pi(\mathbf{X}_{i} | \mathbf{X}_{i-1})\,,
\]
where $\pi(\mathbf{X}_{i} | \mathbf{X}_{i-1})$ is the density of $N_p(\mathbf{X}_{i-1} + f(t, \mathbf{X}_{i-1}; \theta), \Sigma \delta t)$ distribution. 



As an illustrative framework for the spike-and-slab \textit{equation selection} process, suppose we have a first order (stochastic) differential equation as in \eqref{eq:SDE}, where the drift function is a sparse quadratic function
of the state variables and time. Define $\tilde{\mathbf{X}}$ to be the vector containing linear, quadratic, and cross terms of the system components of $\mathbf{X}$, along with linear and quadratic time components, $t$ and $t^2$ appended at the end. Therefore, this vector contains all possible combination of terms up to order 2 that can occur in a dynamic system.  

Let $\tilde{\mathbf{X}}$ then be $p^*$ dimensional and let $B(t,\theta)$ be a $p \times p^*$ matrix so that the drift function for a desired stochastic differential equation can be written as
\[
f(t, \mathbf{X}_{k}, \theta) = B(t,\theta) \tilde{\mathbf{X}}_k\,.
\]
For convenience, we use the notation $B(t,\theta) = B$.
Then, the modified system equation for this generic drift 
function is 
\begin{equation}
\label{eq:SS_drift}
	\dfrac{d \mathbf{X}}{dt} = B\, \tilde{\mathbf{X}} + \Sigma^{1/2} \, dW_{t} \, ,
\end{equation}
with the corresponding Euler-Maruyama discretization being
 \begin{equation*}
  \mathbf{X}_{k+1} = \mathbf{X}_{k} + B\, \tilde{\mathbf{X}}_k\delta t + \sqrt{\Sigma \delta t}\,\xi_{k} \, .
 \end{equation*}
In realistic systems, we expect
many entries of $B$  to be zero,  and consequently many terms in $\tilde{\mathbf{X}}$ 
may not appear in the drift function. 

As an example, 
consider the four-dimensional stochastic Lorenz 96 model with the drift function described in \cite{lorenz:1996}. Here $\mathbf{X}_t = ({X_{t}^1}, \dots, {X_{t}^4})$ and 
\[
	d X_{t}^{i} = [(X_{t}^{i+1} - X_{t}^{i-2}) \, X_{t}^{i-1} - X_{t}^{i} + \theta] \, dt + \Sigma_{i}^{1/2} \, d W_{t}^{i}	\, ,
\]
where $ i \in \{1,2,3,4\}$ are cyclic indices and $\theta \in \mathbb{R}$ is the drift parameter. The drift equations here are often used to represent a simplified atmospheric model \citep[see for example][]{lorenz:eman:1998}. For this system,  $\tilde{\mathbf{X}}$ is
\[
	 [1, X^{1}, X^{2}, X^{3}, X^4, {X^{1}}^{2},  {X^{2}}^{2},  {X^{3}}^{2},  {X^{4}}^{2}, {X^{1}} X^{2}, X^{1} X^{3}, X^{1} X^{4}, X^{2} X^{3}, X^{2} X^{4}, X^{3} X^{4}, t, t^{2}]^{T} \, .
\]
The corresponding $B$ matrix in \eqref{eq:SS_drift} is all 0s, except for non-zero elements in  indices (column-wise) $(1, 2, 3, 4, 5, 10, 15, 20, 39, 42, 44, 46, 52, 53, 55, 57)$  with the first four being $\theta$ and the others being either -1 or 1 appropriately.

This illustrative framework can be easily extended to more complex dynamical systems, including those that potentially involve higher order derivative with respect to time and space variables, higher order polynomials in state, location and time variables, and transformations of such variables.

Continuing with the above illustrative framework, we employ a spike-and-slab prior on the elements of the matrix $B$ to identify the components of $\tilde{\mathbf{X}}$ which have a significant impact on the system underlying the observed data. Our prior choice is based on the works of  \cite{geo:mccul:1993,ish:2005,nari:he:2014}. 
Each component of $B$ is given an independent hierarchical prior as:
\begin{align}
    B_{i,j} | \gamma_{i,j} & \overset{\text{ind}}{\sim} (1 - \gamma_{i,j}) N(0, \tau_{0}^{2}) + \gamma_{i,j} N(0, \tau_{1}^{2}) \label{eq:B_prior}\\
    \gamma_{i,j} & \overset{\text{ind}}{\sim} \text{Bernoulli}(q_{i,j})\, ,
\end{align}
where $\tau_{0} > 0$ is  small and $\tau_{1}$ is at least an order of magnitude larger than $\tau_{0}$. Therefore,
if $\gamma_{i,j} = 0$, the value of $B_{i,j}$ is confined to a narrow region around 0, nullifying its contribution in the model equation. Additional scientific prior knowledge about the parameters $B_{i,j}$'s can be easily absorbed in the $q_{i,j}$'s. 

Additionally, we assume a Gaussian prior on $\mathbf{X}_0$ as well so that $\mathbf{X}_0 \sim N_p(\mu_0, \lambda_0^2)$. We assume that $\Sigma$ is a diagonal matrix with 
$\Sigma_i \overset{\text{iid}}{\sim}  \text{Inverse-Gamma}(\alpha, \beta)$ for $i = 1, \dots, p$. Suppose  $\theta = (\theta_1, \dots,\theta_d) \in \mathbb{R}^{d}$ contains the $d$ drift parameters. The prior on $\theta$ is 
\begin{align*}
	\theta & \sim  N_d(m_0, s^{2}_0 I_d)\, ,
\end{align*}
where $s^{2}_0$ and  $m_0$ are hyper-parameters.
The resulting posterior distribution of $(\theta, \Sigma, \mathbf{X})$ is presented in the supplementary materials. 
An MCMC sampler described in Section~\ref{sec:computation} yields posterior probabilities of association for each $\gamma_{i,j}$. A median cutoff of $0.5$ \citep{rov:george:2018} on the posterior probabilities of association is employed to determine whether a component is to be deemed active or inactive. 


Once a list of final important components is determined by the above process and a reduced stochastic differential equation structure identified using the above process, we propose a second round of computation using only the reduced equation system to obtain posterior distributions of the finite dimensional parameters.  We are fully aware of the theoretical and statistical difficulties arising from such two-stage computations, and  the super-efficient nature of the estimation and inference obtained from the second round of computations. While  caution about superefficiency is valid, the reason for our advocacy of a two-stage computational process is more practical: the initial model 
in \eqref{eq:SS_drift} may be very high-dimensional and the posterior distributions of the true (non-zero) parameters too imprecise for scientific use, especially considering the fact that we have limited and sparse data at hand. We use the term \textit{inference model} for the reduced system discovered by the above Bayesian equation selection process.

\section{Posterior Computation}
\label{sec:computation}

The  posterior density for the spike-and-slab model is given by
\begin{align*}
	\pi(\mathbf{X}, B, \Sigma, \gamma \mid \mathbf{Y}) & \propto \pi(\mathbf{Y} \mid \mathbf{X}, B ) \, \pi(\mathbf{X} \mid B, \Sigma) \, \pi(B \mid \gamma) \, \pi(\Sigma)\,,
\end{align*}
and is naturally intractable, with MCMC methods being used to obtain posterior estimates. There are some obvious challenges here. First, the overall dimension of $\mathbf{X}$ is typically quite large since $\delta t$ is usually small. In addition, there is high correlation across time in the $\mathbf{X}$s, implying that component-wise updates on the $\mathbf{X}$s would lead to debilitating performance. Thus, we will focus on full block updates for the $\mathbf{X}$s.

\begin{figure}[htbp]
	\centering
	\includegraphics[width = 4in]{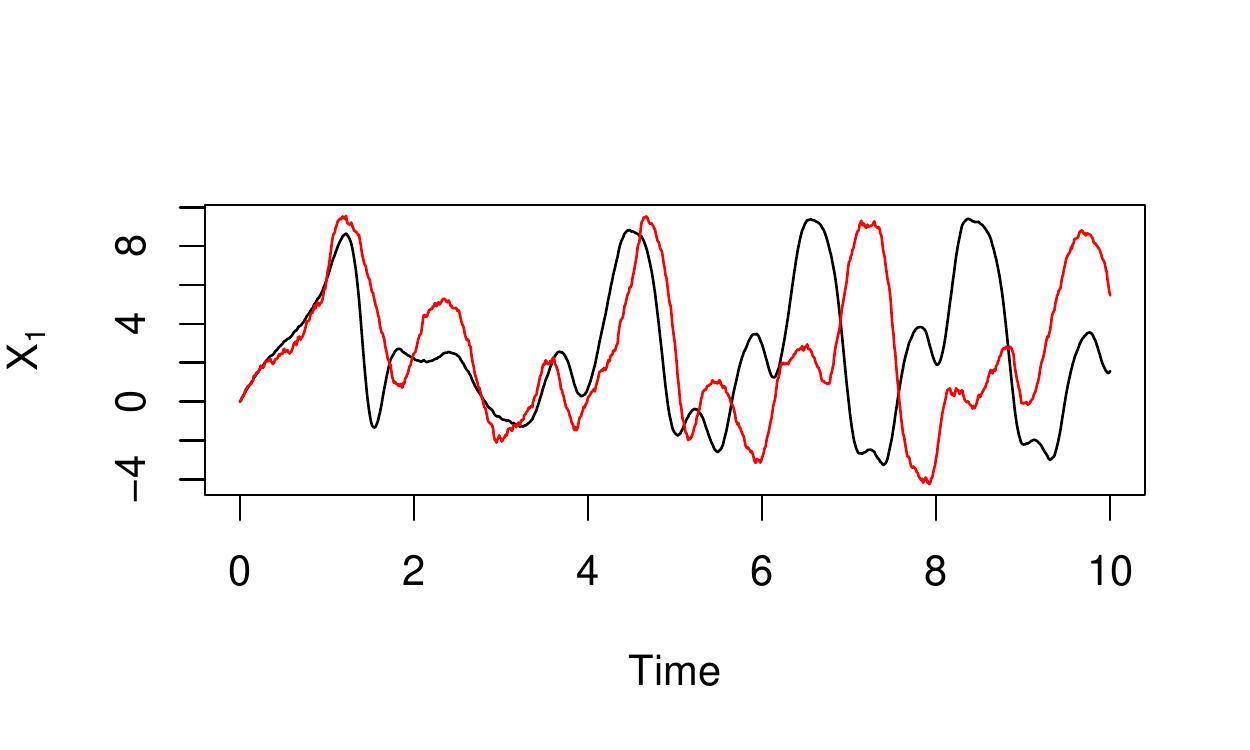}
	\caption{Two realizations of the first component of Lorenz 96 with $\Sigma_1 = .05$ (black) and $\Sigma_1 = 1$ (red).}
	\label{fig:96_sigma_affect}
\end{figure}
There is a particularly delicate relationship between $\Sigma$ and $\mathbf{X}$. Small changes in $\Sigma$ leads to a butterfly effect in the trajectory of the system. Consider Figure~\ref{fig:96_sigma_affect} where we present two trajectories of the first component of a Lorenz 96 system with two different values of $\Sigma$ over 10 time units. In the beginning, the two processes behave similarly, but with time the difference in the trajectories is exaggerated. This implies that small shifts in the estimation of the $\mathbf{X}$s yield large deviations in the $\Sigma$s and vice-versa. Additionally, this subtle balance restricts the movement of any Markov chain, encouraging tiny steps and weak exploration. This was also noticed in the Hamiltonian Monte Carlo algorithm implemented in \cite{vrettas:cornford:2011}. 

We construct a \textit{linchpin} sampler presented in \cite{archila:2016} where we integrate out $\Sigma$ and run an MCMC kernel that keeps the marginal posterior of $\pi(\mathbf{X}, B, \gamma | \mathbf{Y})$ stationary. The draws from this marginal are then cycled through to the full conditional distribution of $\Sigma$, allowing its estimation as well. Notice that,
%
\[
\pi(\mathbf{X}, B, \Sigma, \gamma \mid \mathbf{Y}) = \pi(\Sigma \mid \mathbf{X}, B, \gamma, \mathbf{Y}) \, \pi(\mathbf{X}, B, \gamma \mid \mathbf{Y})\,,
\]
with each diagonal $i$ of $\Sigma$ having the full conditional,
\begin{equation}
	\label{eq:sigma_ss_full_conditional}
	\Sigma_{i} | \mathbf{X}, B, \gamma, \mathbf{Y} \sim \text{inv-gamma} \left(\dfrac{N}{2} + \alpha, \, \beta + \dfrac{\delta t}{2}\, \sum_{j=0}^{N-1} \left(\dfrac{\delta \mathbf{X}_{j+1}}{\delta t} - B \tilde{\mathbf{X}}_{j}\right)^{2}_{i}\right)\,,
\end{equation}
where $(\cdot)_i$ represents the $i$th element of the vector. The full conditional of each $\gamma$ is
\begin{equation}
	P(\gamma_{i,j} = 1 | \textbf{X}, \theta, \textbf{Y}) = \dfrac{\left(\dfrac{q_{i,j}}{\sqrt{2 \pi \tau_{1}^{2}}} \exp \left(- \dfrac{B_{i,j}^{2}}{2 \tau_{1}^{2}}\right)\right)}
	{\dfrac{q_{i,j}}{\sqrt{2 \pi \tau_{1}^{2}}} \exp \left(- \dfrac{B_{i,j}^{2}}{2 \tau_{1}^{2}}\right) + \dfrac{1-q_{i,j}}{\sqrt{2 \pi \tau_{0}^{2}}} \exp \left(- \dfrac{B_{i,j}^{2}}{2 \tau_{0}^{2}}\right)} \,.
\end{equation}
 Exact calculations and expressions can be found in the supplementary materials.
Since the marginal posterior distribution of $(\mathbf{X}, B, \gamma)$ demonstrates significantly less correlation in its components, the linchpin sampler moves with considerable freedom and mixes much faster. 

For updating $(\mathbf{X}, B, \gamma)$, we employ a component-wise MCMC algorithm with random-walk Metropolis-Hastings (MH) updates for $\mathbf{X}$ and each element of $B$ and a Gibbs update on $\gamma$. Thus the number of components in the MCMC update are $2pp^* + 1$. Before zeroing in on the random-walk MH update, we tested the  Hamiltonian Monte Carlo and the No-U-Turn-Samplers. Since the posterior distribution, particularly for the $\mathbf{X}$s lies in a narrow region, we found that both algorithms via their implementation in Stan, struggled to either reach or stay in this region. In addition, the leapfrog integrator required a large number of steps, leading to an implementation that was significantly slower than the random-walk MH updates available in the \texttt{R} package \texttt{mcmc} \citep{geyer:john:2020}. 
Similar struggles are witnessed in the random-walk MH update for $\mathbf{X}$ as well, however, with some careful tuning of the scaling, the random-walk MH significantly improves the mixing in the sampler. 

For the inference model (i.e., the model obtained by the Bayesian equation selection process of the previous section), a similar linchpin sampler is possible where
\[
\pi(\mathbf{X}, \theta, \Sigma | \mathbf{Y}) = \pi(\Sigma | \mathbf{X}, \theta, \mathbf{Y}) \, \pi(\mathbf{X}, \theta| \mathbf{Y})\,,
\]
with each diagonal $i$ of $\Sigma$ having the full conditional
\begin{equation}
	\label{eq:sigma_full_conditional_inf}
	\Sigma_{i} | \mathbf{X}, \theta, \mathbf{Y} \sim \text{inv-gamma} \left(\dfrac{N}{2} + \alpha, \, \beta + \dfrac{\delta t}{2}\, \sum_{j=0}^{N-1} \left(\dfrac{\delta \mathbf{X}_{j+1}}{\delta t} - f(t, \mathbf{X}_{j}, \theta)\right)^{2}_{i}\right)\,.
\end{equation}
The MCMC algorithm employed for the marginal posterior of $(\mathbf{X}, \theta)$ is a random-walk MH with scaling chosen to obtain around 23\% acceptance, as recommended by \cite{rob:gel:gilks}.

The slow mixing of the Markov chains in such high-dimensions brings a critical challenge of choosing starting values for the Markov chain. Since $\Sigma$ is integrated out, starting values of $\Sigma$ are not required. However, choosing starting values of the $\mathbf{X}$s is particularly challenging. The joint posterior distribution of the $\mathbf{X}$s takes mass in a narrow region, from where it is close to impossible to start the Markov chain unless the true latent process is known. As also discussed in \cite{soren:2004}, a reasonable starting value is to interpolate the observed $\mathbf{Y}$s, which is what we employ here. Due to the narrow high probability region, one cannot afford to make large jumps for the $\mathbf{X}$s, implying slow converge of the latent process. Naturally, this in turn also affects the $\Sigma$s. However, we note that, in general, the drift parameters, $\theta$ do not contribute significantly to the slow mixing rate.

\section{Examples}
\label{sec:examples}

We implement our proposed spike-and-slab model for system identification for data generated from three systems: 
(i) Lorenz 96 (L96) (ii) Ornstein-Uhlenbeck system (OU) (iii) Lorenz 63 (L63)\footnote{The code is available at \texttt{https://github.com/kushagragpt99/BEqSelection}}. The data generation settings for each 
system are in Table~\ref{table:data_settings}. For the hyper-parameter settings for this model, we use the rule of 
\cite{nari:he:2014} multiplied by an appropriate scaling factor so that $\tau_{0}$ and $\tau_{1}$ lie in the range $(0,1)$ and $(2,10)$, respectively. These values ensure that there is adequate separation between the spike and slab components to deter random jumping of $\gamma$ values while allowing for easy switching in $\gamma$ if the model finds corresponding components to be significant (or insignificant).

\begin{table}[htbp]
	\caption{Parameter settings for the examples. Additionally, $\delta t = .01$, $N_{\text{obs}} = 20$ and $R = .05$ for all systems.}
	\label{table:data_settings}
	\begin{center}
		\begin{tabular}{|c|c|c|c|c|c|c|c|c|c|c|}
		\hline
			System & $t_K$ &  $\theta$ & $\Sigma$ & burn-in & \# parameters & $\tau_{0}$ & $\tau_{1}$ \\ 
			\hline
			L96 & 10  & 8 & 0.5 & 50 & 4,009 & 0.13 & 4.52\\ 
			L63 & 20 & (10, 28, 8/3) & 0.6  & 5000 & 6,009 & 0.50 & 5.00\\ 
			OU & 2 &  2 & 1.0 & 50 & 203 & 0.09 & 2.90\\ \hline
		\end{tabular}
	\end{center}
\end{table}

Practitioners typically have prior information about which system may best align with their data. Thus, we set $q_{i,j} = .90$ for the elements of $B$ which are deemed active under this prior knowledge and set $q_{i,j} = .10$ for elements of $B$ which are deemed inactive. Once a system is identified, we then run the Bayesian inferential model to obtain posterior estimates of $\theta$, $\Sigma$, and $\mathbf{X}$. 

\subsection{Lorenz 96}
\label{subsec:l96}
Recall the four-dimensional stochastic Lorenz 96 model introduced in Section~\ref{sec:bayesian_model}:
\[
	d X_{t}^{i} = [(X_{t}^{i+1} - X_{t}^{i-2}) \, X_{t}^{i-1} - X_{t}^{i} + \theta] \, dt + \Sigma_{i}^{1/2} \, d W_{t}^{i}	\, ,
\]
where $ i \in \{1,2,3,4\}$ are cyclic indices and $\theta \in \mathbb{R}$ is the drift parameter.  

For generating a Lorenz-96 trajectory using Euler-Maruyama approximation, we use $\delta t = 0.01$ for the time interval $T 
= [0,10]$ with the true value of the parameter as $\theta = 8$. We have an additional burn-in of $T = 50$ 
time units to allow the process to reach a level of stationarity. Due to the Euler-Maruyama discretization, $N = 1000$ values are generated after this initial burn-in. We observe 20 data points per time unit, so that overall $K = 200$. This takes the total number of estimated 
parameters ($\textbf{X}, \theta$) and the diagonal elements  of $\Sigma$) to $4,009$ ($4 \times 1001 + 1 + 4$). 
The true diagonal covariance matrix $\Sigma$ is $0.5 \, I_{4}$ and we fix $R = 0.05 \, I_{4}$.

\begin{figure}[htbp]
	\centering
	\includegraphics[width = 2.8in]{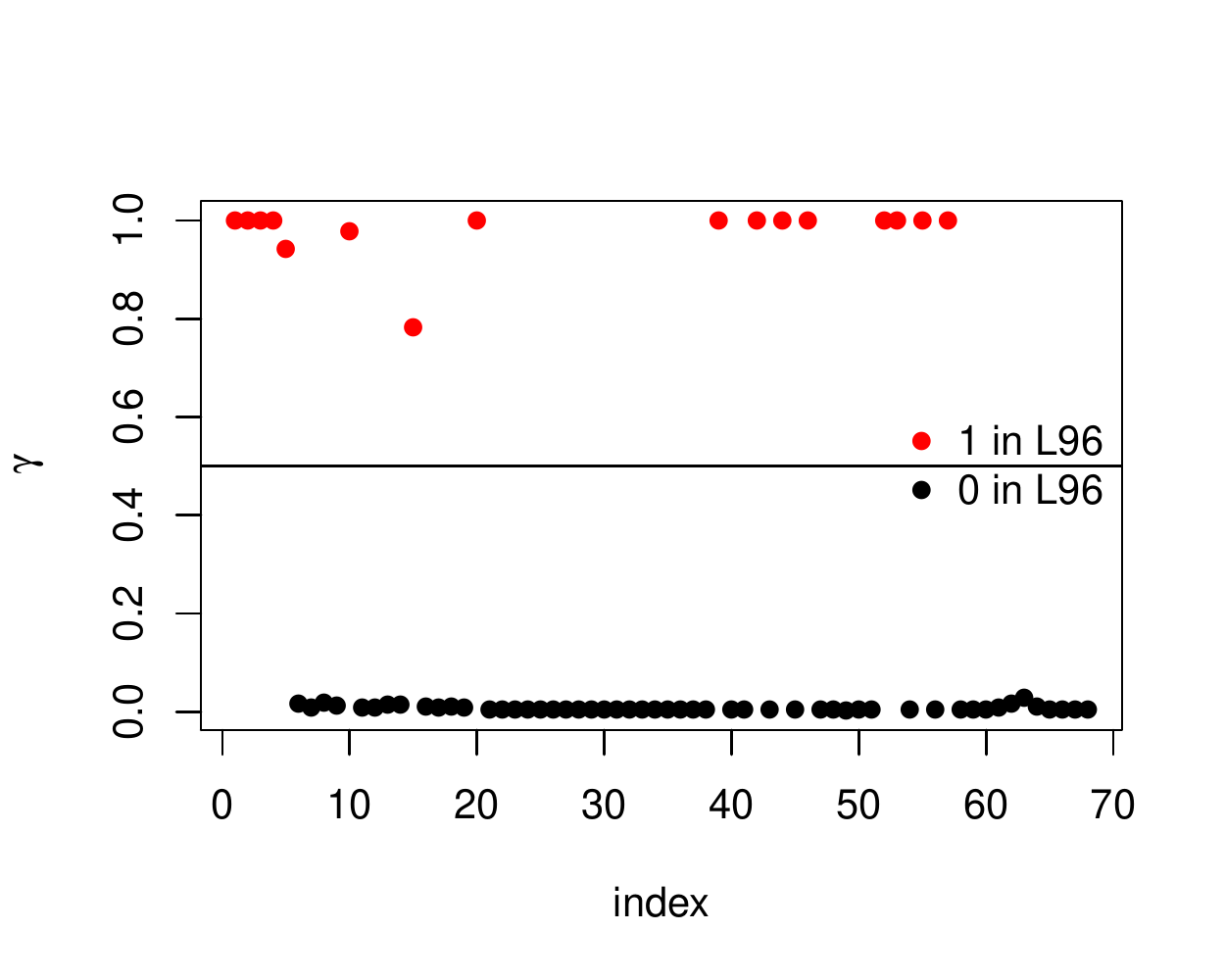}
	\includegraphics[width = 2.8in]{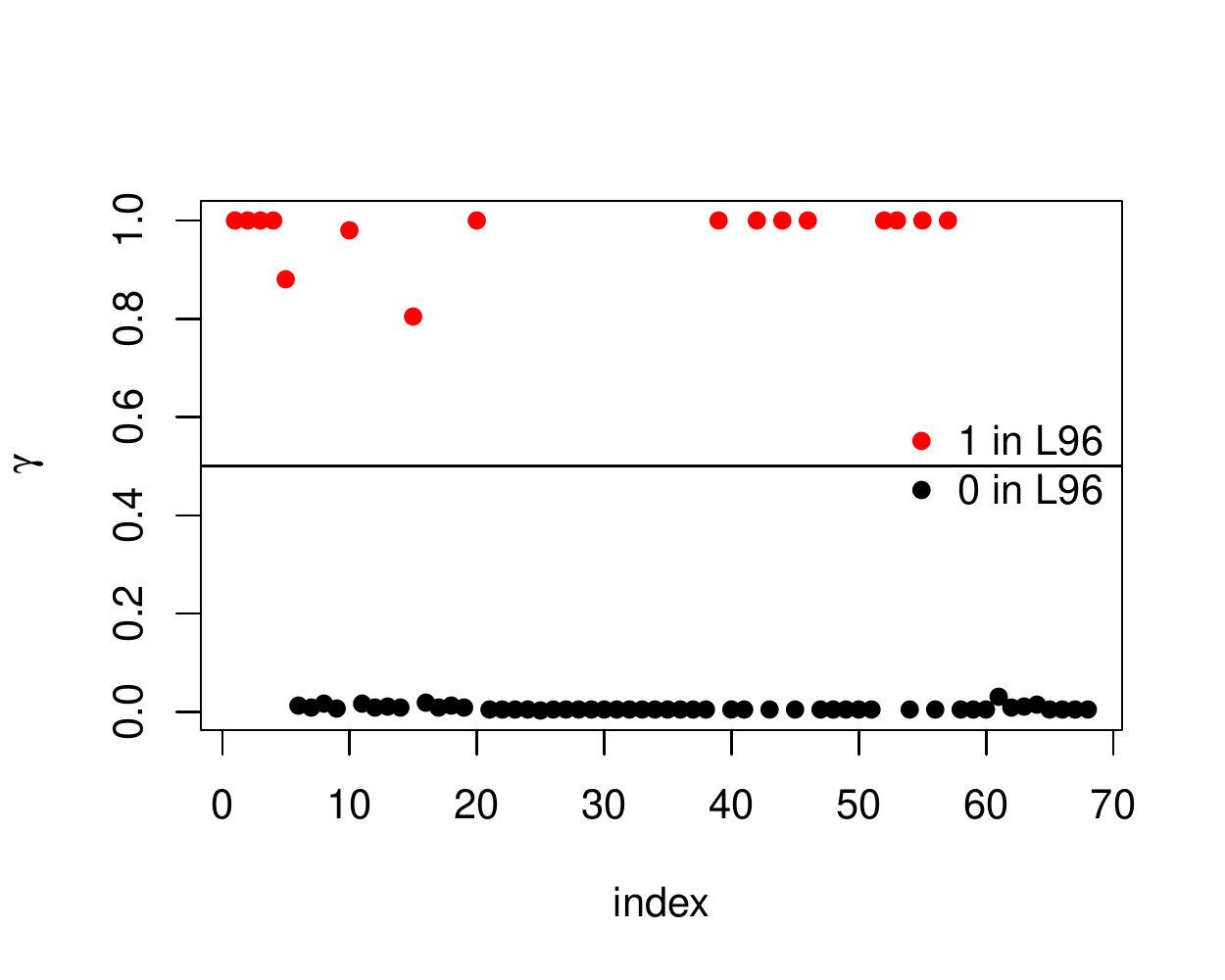}
	\caption{Posterior mean estimates of $\gamma$ for each element of $B$ for Lorenz 96, with \textbf{X} initialized at its truth (left) and interpolated starting values (right).}
\label{fig:gamma_lorenz96}	
\end{figure}

First, we perform model identification for Lorenz-96 using our proposed spike-and-slab model. We run the linchpin 
samplers initialized both at the truth and with interpolated values of $\mathbf{X}$ for Monte Carlo sample sizes 
50000 and $10^5$, respectively. We mention here that longer Markov chain lengths do not impact 
the inference on the $\gamma$s, thus we limit the MCMC to shorter runs for this step. In Figure~\ref{fig:gamma_lorenz96}, we present posterior mean estimates of the $\gamma$s, color coded with what their value should be under the Lorenz-96 model. Red implies active, and black implies inactive. We find that the quality of inference is the same for both runs and the Lorenz 96 system is correctly identified.

Having identified the Lorenz-96 model, we run the inference model for $10^{6}$ steps using the Lorenz-96 drift function in order to estimate parameter, $\theta$, and infer the $\Sigma$s.
We set $m_0$ to be centered at the estimate of the parameter from the spike-and-slab estimates, in order to allow some reasonable centering in the prior information for this chaotic system.

\begin{figure}[htbp]
	\centering
		\includegraphics[width = 2.2in]{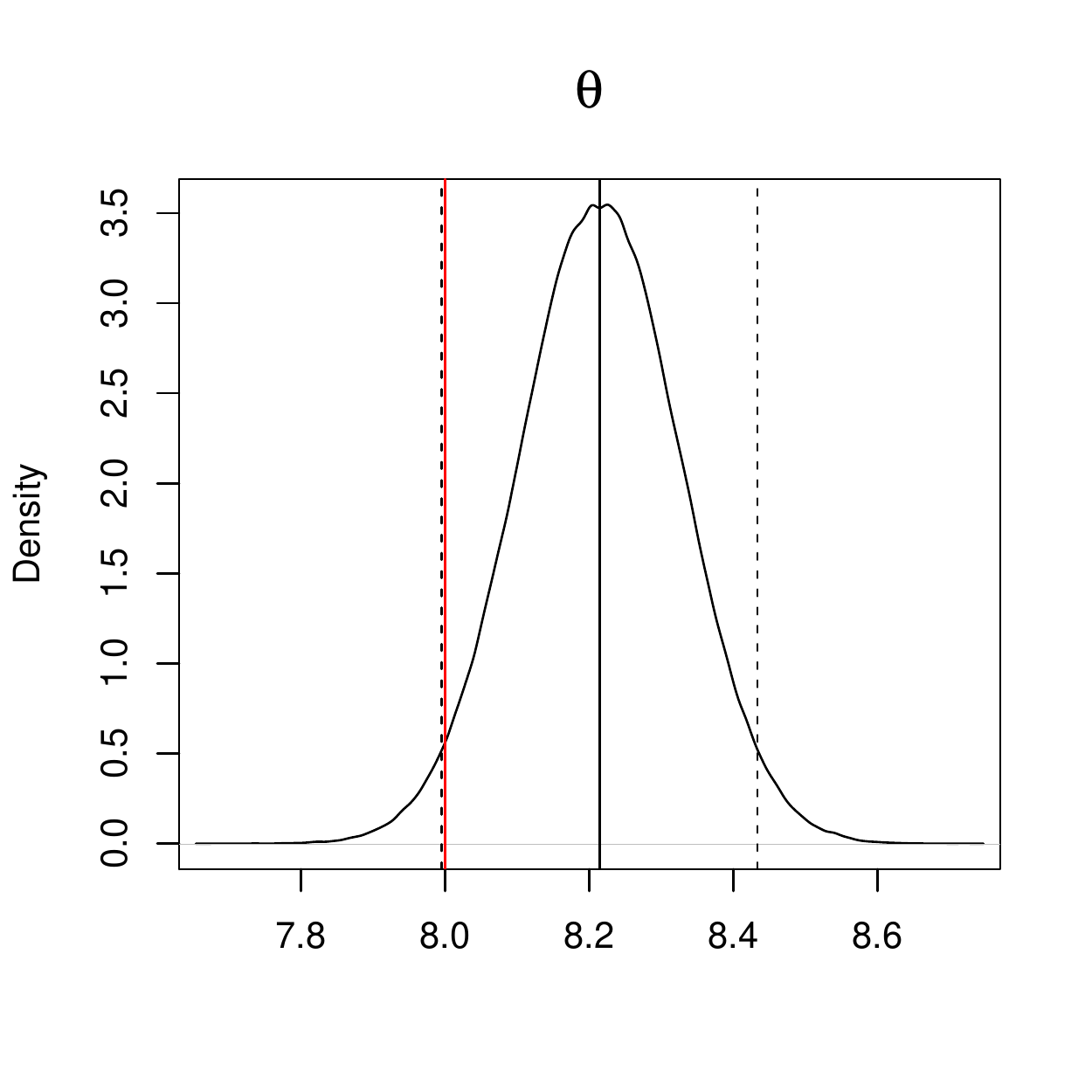}
		\includegraphics[width = 2.2in]{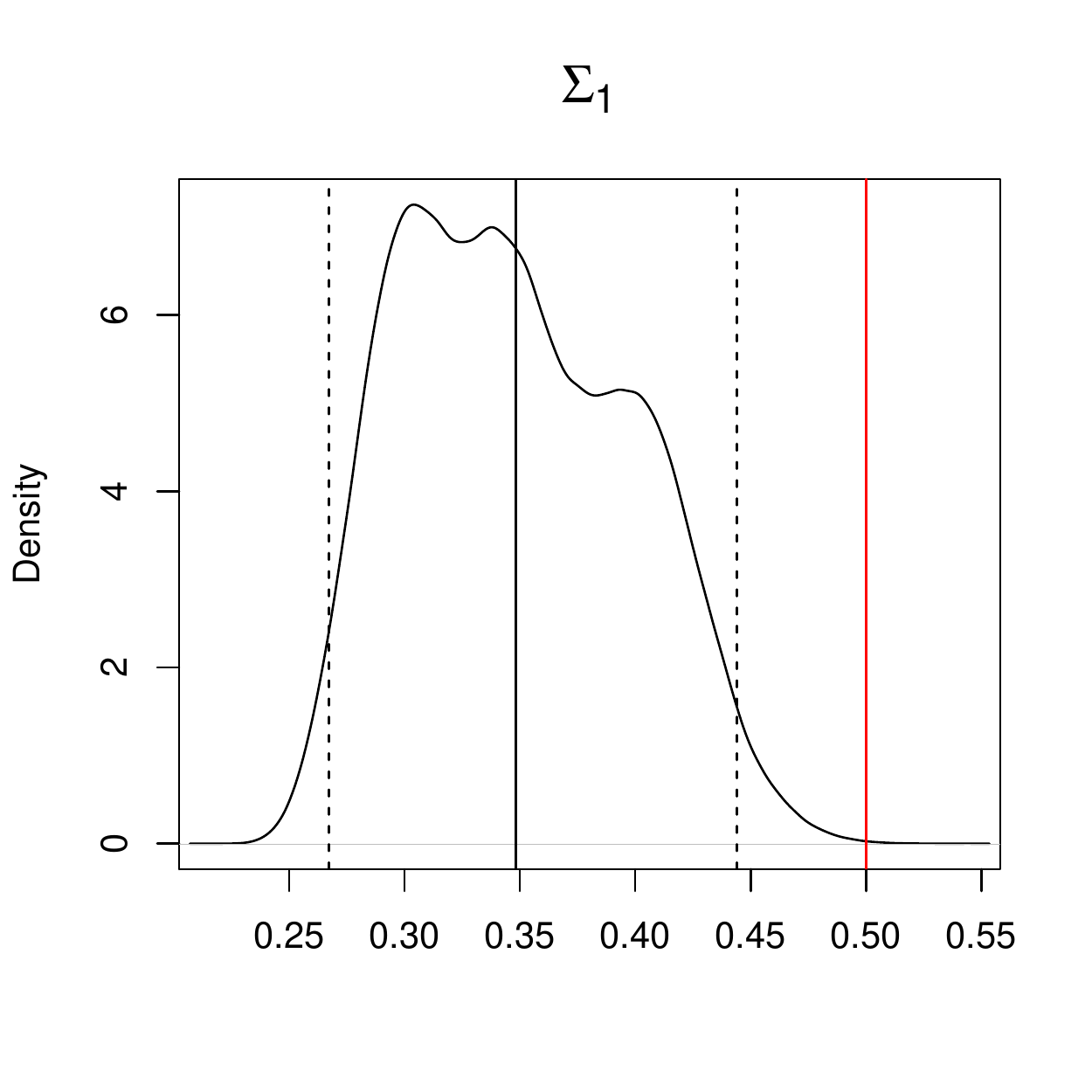}	
		\includegraphics[width = 2.2in]{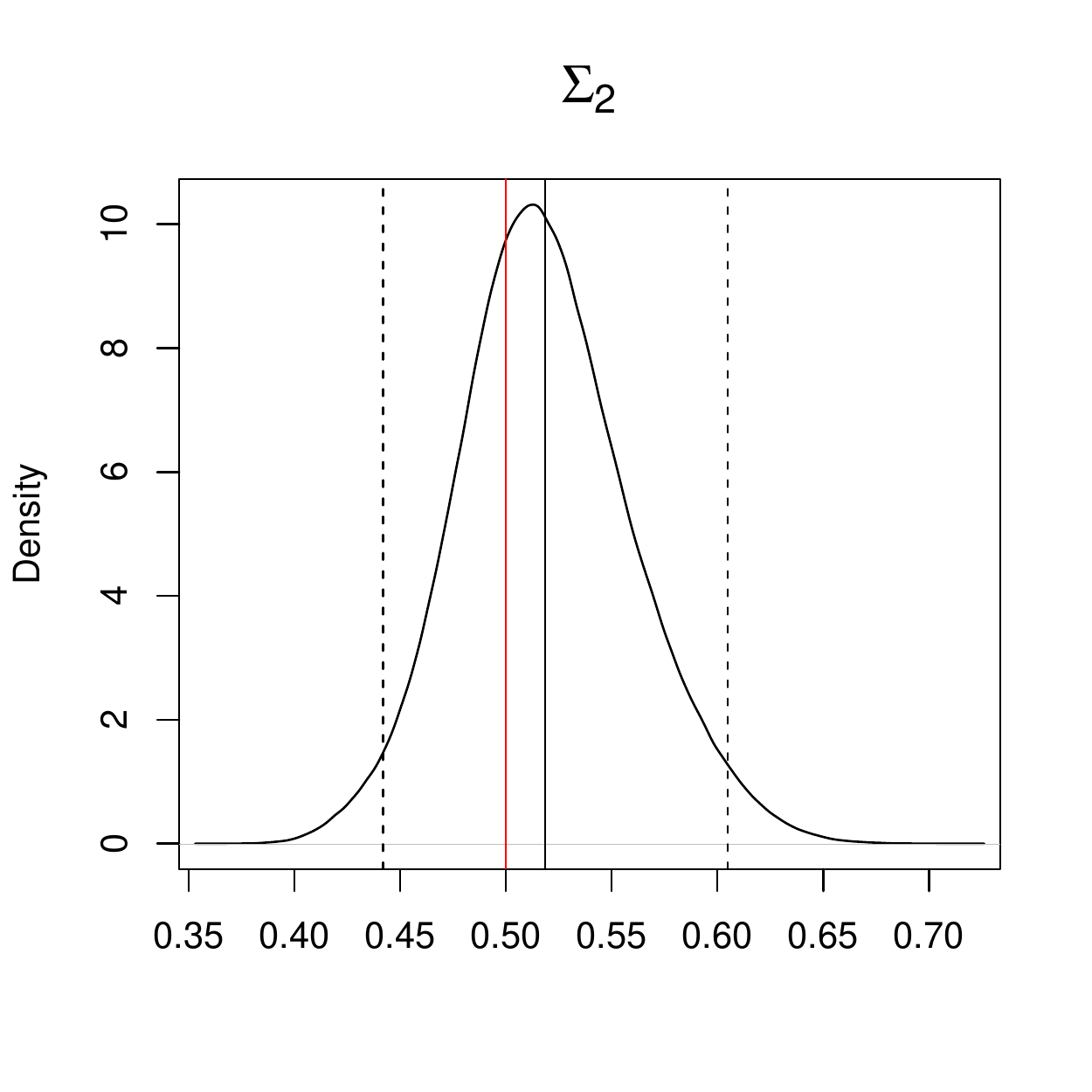}
		\includegraphics[width = 2.2in]{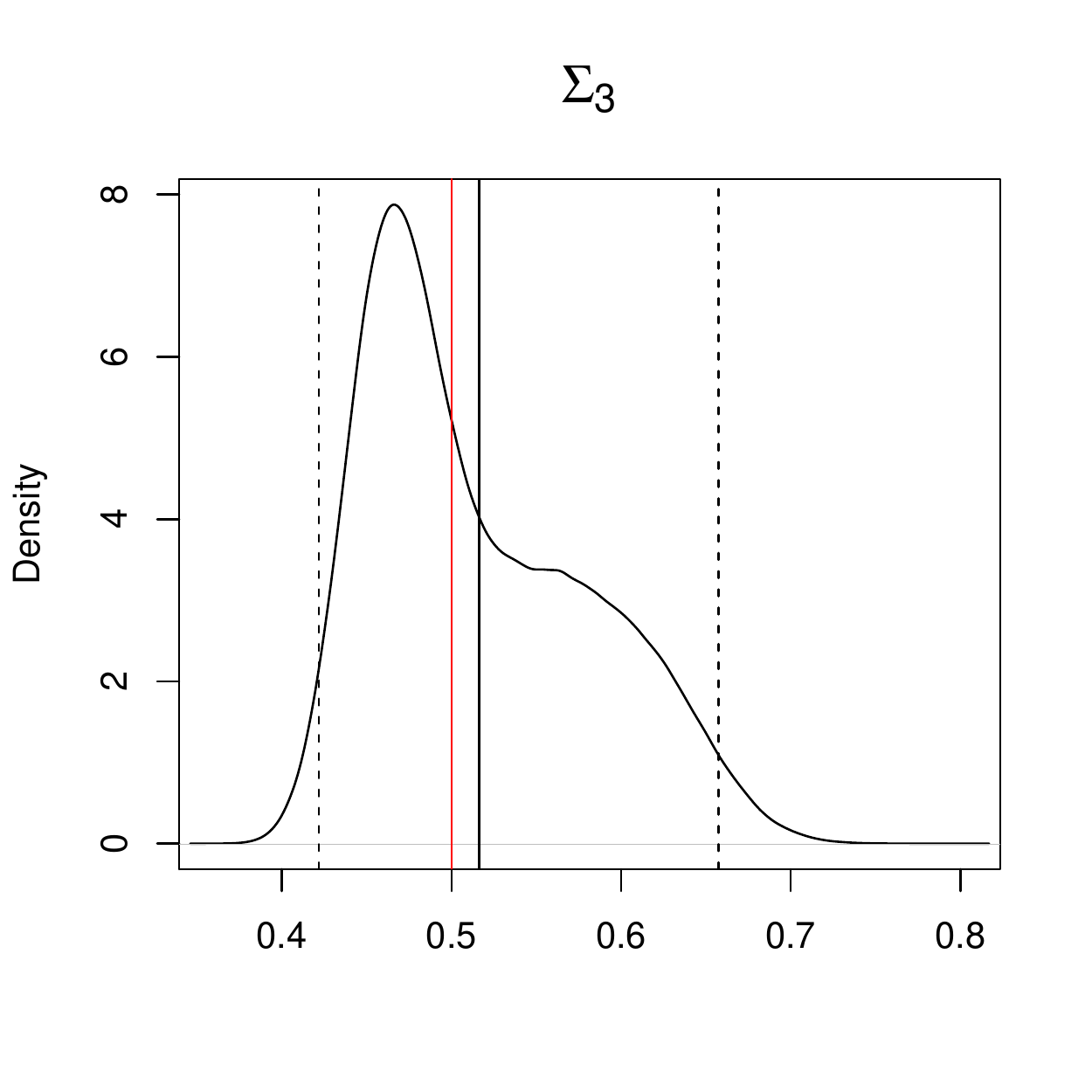}	
		\includegraphics[width = 2.2in]{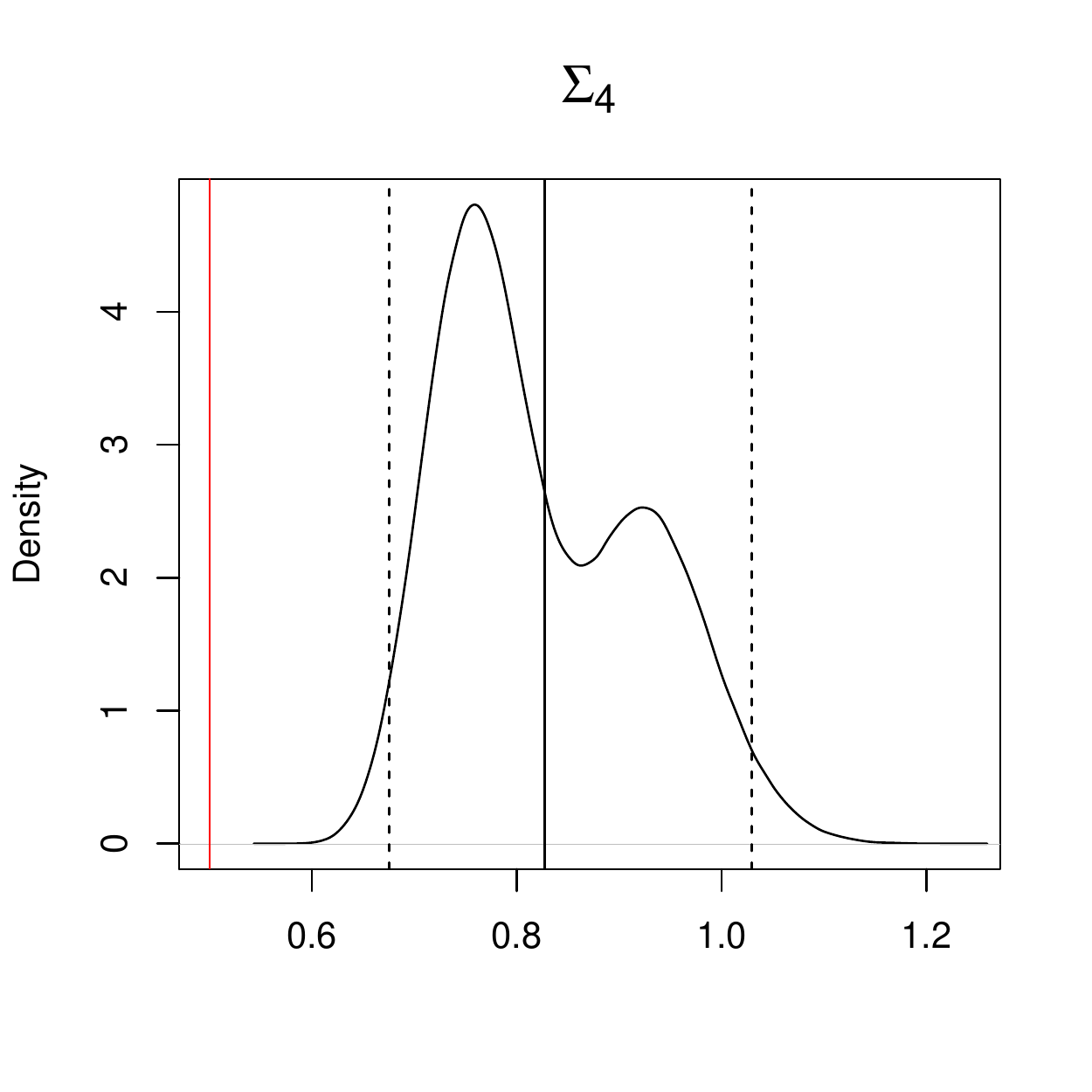}	
		\caption{Posterior density estimates of $\theta$ and the diagonals of $\Sigma$ for Lorenz 96. Results are for interpolated $\mathbf{X}$ as starting values.}
		\label{fig:l96_interp}	
	\end{figure}

We present results for the run from the interpolated starting point of the $\mathbf{X}$s. Due to the slow mixing of 
the inferential model, we ran the Markov chain for 5 million steps and noticed that for the first 4 million, the 
Markov chain demonstrated a lack of stationarity and a significant drift in the estimates of $\Sigma$. Thus only the last 
1 million steps are used for estimation and their posterior density estimates are presented in Figure~\ref{fig:l96_interp}. 
It is clear that the parameter $\theta$ is well estimated. However, as is expected from the vulnerable relationship 
between the $\mathbf{X}$s and $\Sigma$, the $\Sigma$s are reasonably estimated, although some discrepancy remains due to lack 
of information on the true $\mathbf{X}$s. 

\begin{figure}[htbp]
	\centering
		\includegraphics[width = 3in]{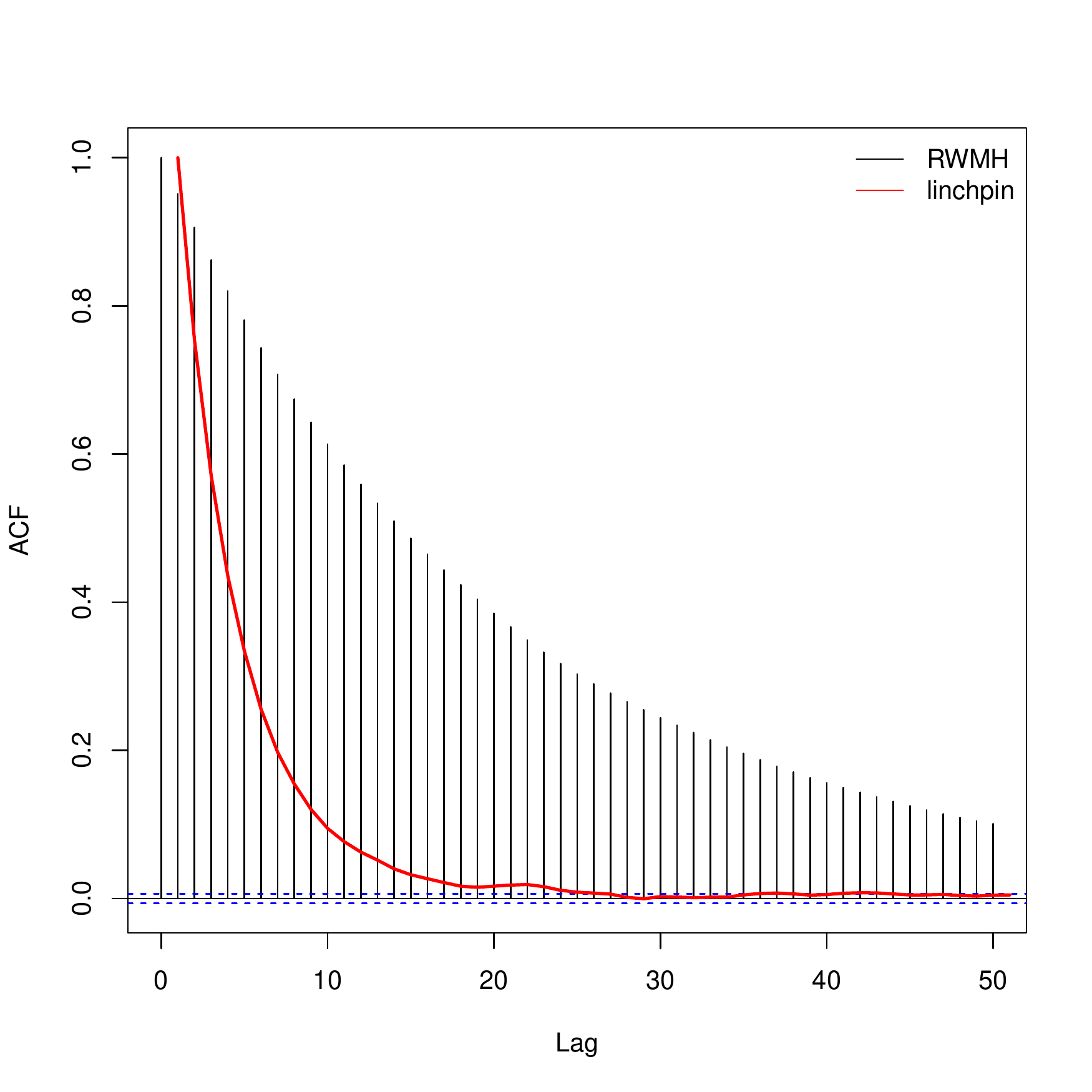}
		\caption{Autocorrelation of vanilla MH vs linchpin sampler for the inference model.}
\label{fig:l96_rwmh_vs_linch}		
\end{figure}

A comparison of the linchpin sampler and vanilla MH on $(\theta, \Sigma, \mathbf{X})$ shows 
that the Markov chain for the former displays significant improvements over vanilla MH. Figure \ref{fig:l96_rwmh_vs_linch} shows 
the ACF for the drift parameter plots of the two samplers, with correlation in the chain decaying at a faster rate
for linchpin sampler than vanilla MH. Additionally, the effective sample size (calculated using \texttt{R} package \texttt{mcmcse} \citep{mcmcse}) for $\theta$ for the two samplers is $12697$
and $2688$ for a $10^{5}$ length run, respectively, indicating the superiority in performance of our linchpin sampler.

\begin{figure}[htbp]
	\centering
		\includegraphics[width = 3in]{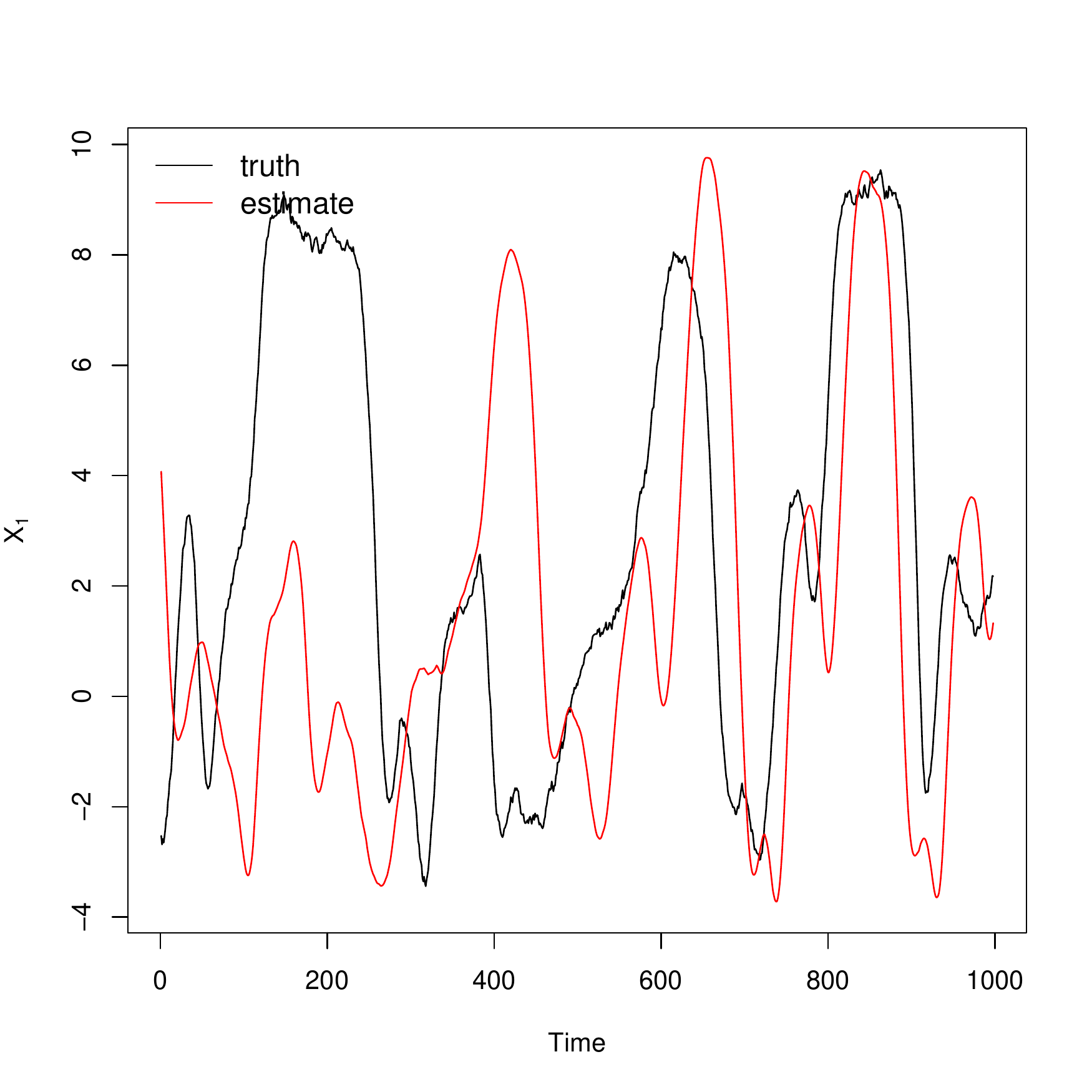}
		\caption{Truth vs posterior mean trajectories of the first component of Lorenz-96.}
		\label{fig:l96_traj}
\end{figure}

\subsection{Ornstein-Uhlenbeck}

The Ornstein–Uhlenbeck (OU) is a univariate process defined by the following SDE 
\[
	d x_{t} = - \theta x_{t} dt + \Sigma d W_{t} \,,
\]
where $\theta \in \mathbb{R}$ is the drift parameter.
The OU trajectory is generated for the time interval $T = [0,2]$ with the drift parameter $\theta = 2$ using the 
Euler-Maruyama approximation with $\delta t = 0.01$ discretization steps. Similar to Lorenz-96, we have a burn-in of 
$T = 50$ time units for the EM approximation for the model to reach stationarity. The total number of parameters to be
estimated ($\textbf{X}, \theta$ and  $\Sigma$) are $203$ ($201 + 1 + 1$), with $K = 40$
observations. The observation error is fixed at $R = 0.05$, and we use $\Sigma = 1$ to generate the process.

\begin{figure}[htbp]
	\centering
		\includegraphics[width = 2.2in]{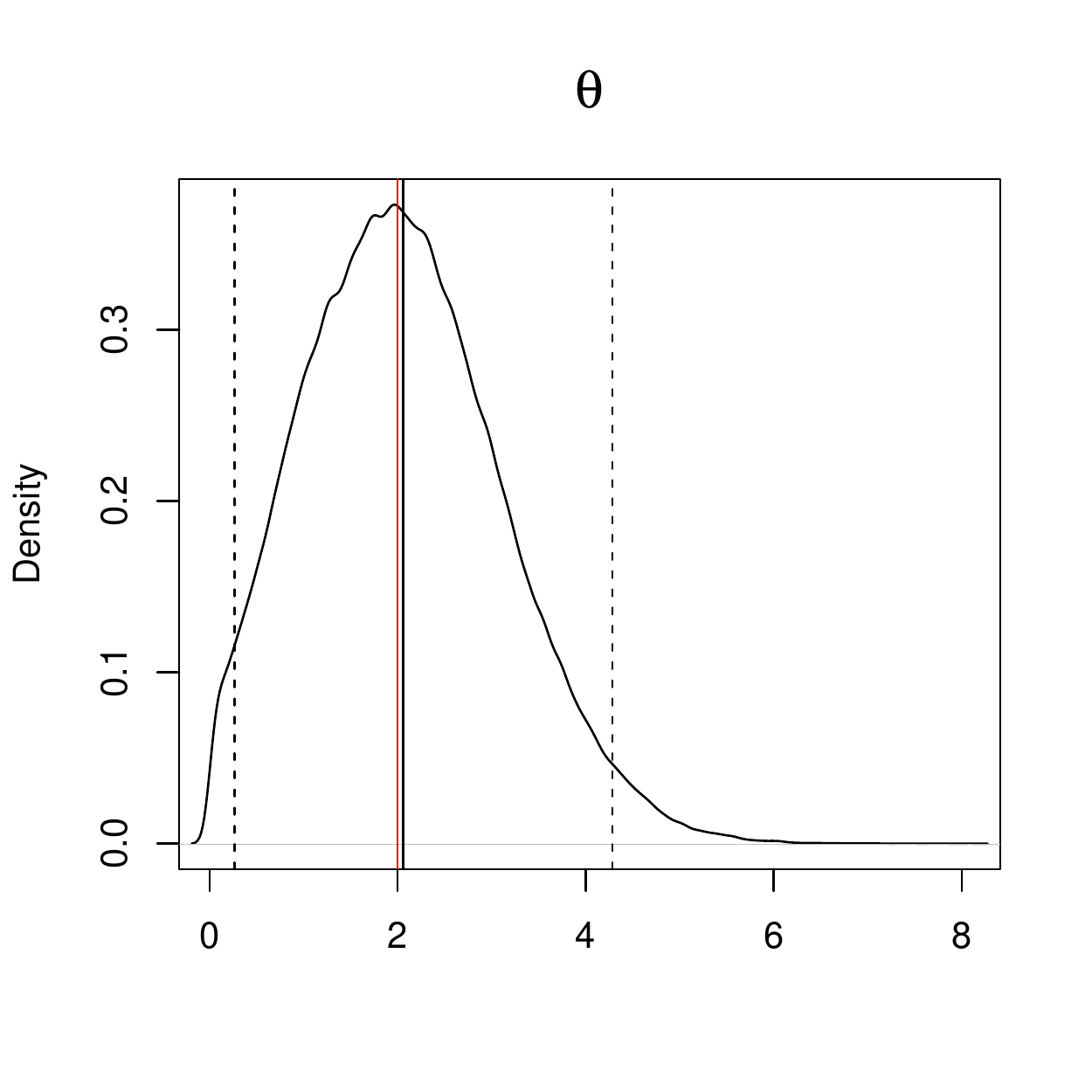}
		\includegraphics[width = 2.2in]{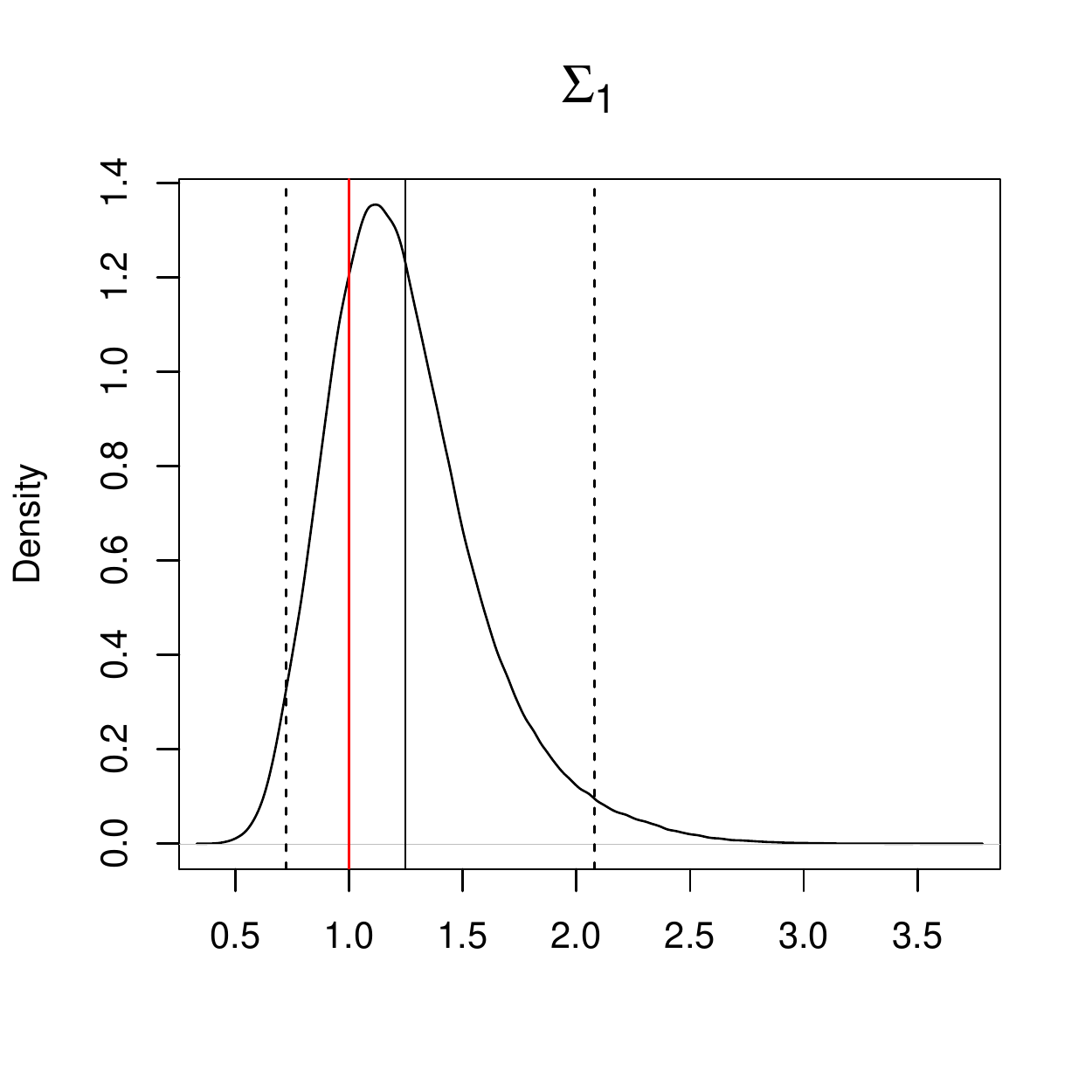}		
		\caption{OU estimated marginal posterior densities. Results are for interpolated $\mathbf{X}$ as starting values.}
		\label{fig:OU_interp_inf}		
\end{figure}
\begin{figure}[H]
	\centering
	\includegraphics[width = 2.8in]{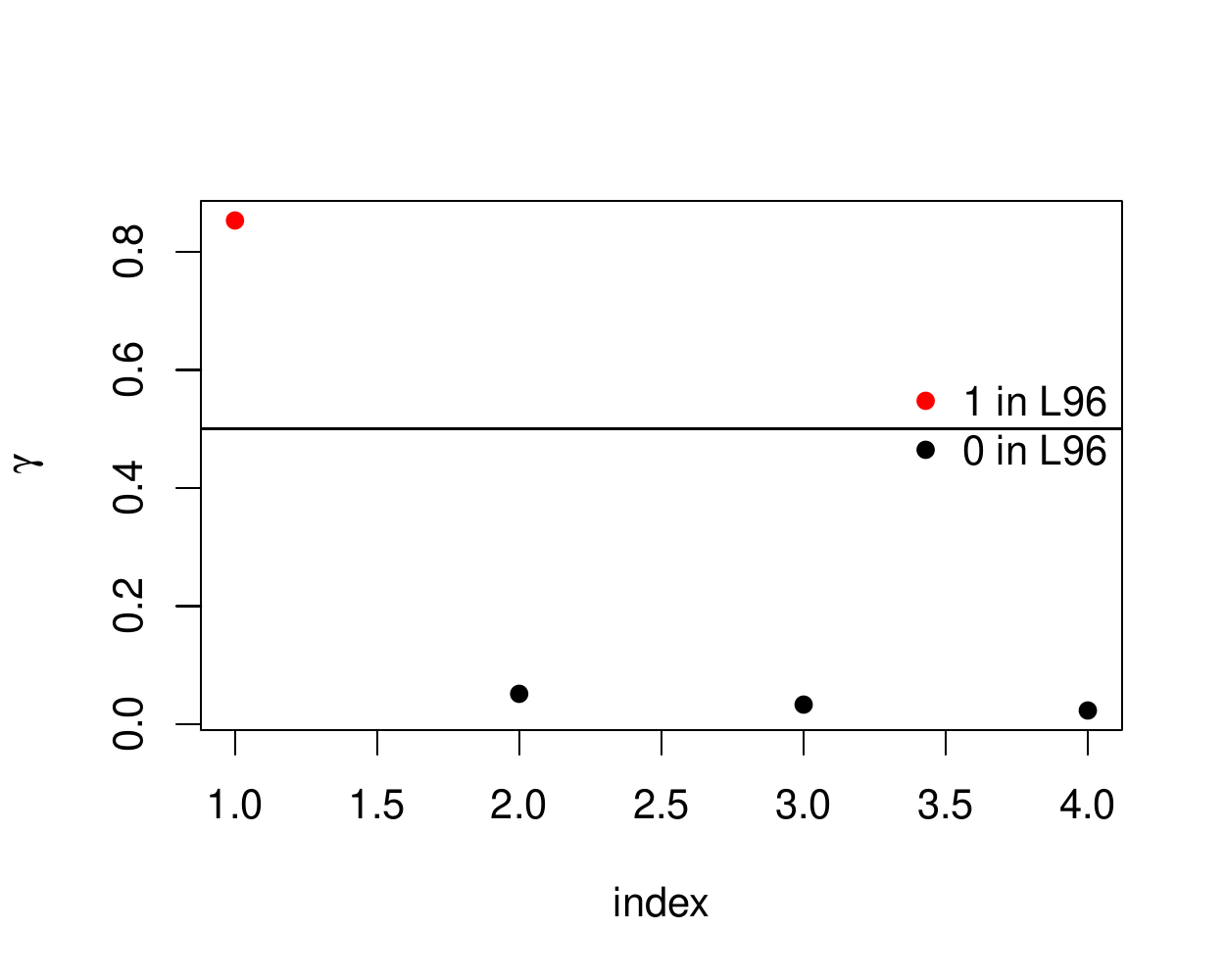}
	\includegraphics[width = 2.8in]{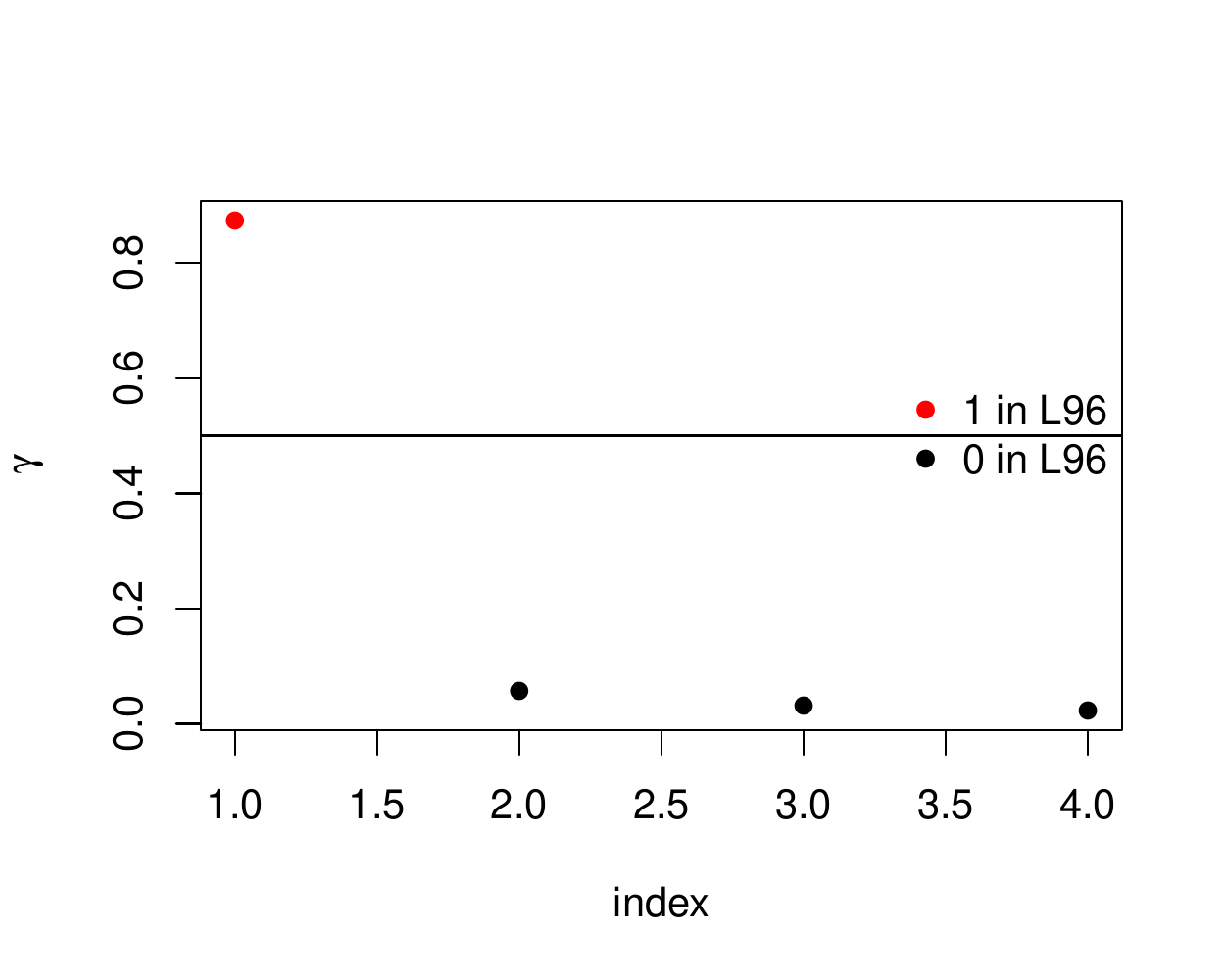}
	\caption{Posterior mean estimates of $\gamma$ for OU, with $\textbf{X}$ initialized at its truth (left) and interpolation (right). }
\label{fig:OU_gamma}	
\end{figure}

We first perform model identification for the OU system using samplers initialized at the truth and at interpolations
from \textbf{Y}, using run lengths of $50000$ and $10^{5}$ respectively. The results from our spike-and-slab model  in Figure~\ref{fig:OU_gamma} show that the OU system is correctly identified. Next, we run the linchpin based inference
model for $10^{6}$ steps with the OU drift function for \textbf{X} initialized at values interpolated from \textbf{Y}. Similar 
to Lorenz-96, we set $m_{0}$ as the estimate of the parameter from the spike-and-slab model. Figure \ref{fig:OU_interp_inf}
shows that $\theta$ and $\Sigma$ are estimated well. Since OU had only $200$ latent variables to estimate, compared to 
$4004$ for Lorenz-96, the sampler has a much easier job in estimation, the effect of which is translated into better quality
estimates of $\theta$ and $\Sigma$.



\subsection{Lorenz-63}

The stochastic Lorenz-63 is driven by the following SDE:
\begin{align}
\label{eq:lorenz63}
d\mathbf{X}_{t} &= \begin{bmatrix}
                    \sigma \left(y_{t} - x_{t}\right) \\
                    \rho x_{t} - y_{t} - x_{t}z_{t} \\
                    x_{t}y_{t} - \beta z_{t}
                    \end{bmatrix} dt + \Sigma^{1/2}d\textbf{W}_{t} \,,
\end{align}

where $(\sigma, \rho, \beta) \in \mathbb{R}^{3}$ are the drift parameters.   

We use Euler-Maruyama approximation with $\delta t = 0.01$ discretization to generate Lorenz-63 trajectory 
for the time interval $T = [0,20]$. We use parameters $\theta = [\sigma, \rho, \beta] = [10, 28, 8/3]$ since this choice exhibits  
chaotic behavior. Similar to the previous two examples, an additional burn-in of $T = 5000$ 
time units for the EM approximation allow the chaotic model to reach stationarity. We generate $N = 2000$ values 
after the burn-in, taking the total number of estimated 
parameters ($\textbf{X}, \theta$ and the diagonal elements  of $\Sigma$) to $6,009$ ($3 \times 2001 + 3 + 3$). 
The diagonal covariance matrix $\Sigma$ is $0.6 \, I_{3}$ and the covariance matrix of the observation matrix 
$R = 0.05 \, I_{3}$. We observe $20$ values of \textbf{X} per time unit with added measurement error, giving a 
total of $K = 400$ real observations \textbf{Y}.

\begin{figure}[htbp]
	\centering
 	\includegraphics[width = 2.8in]{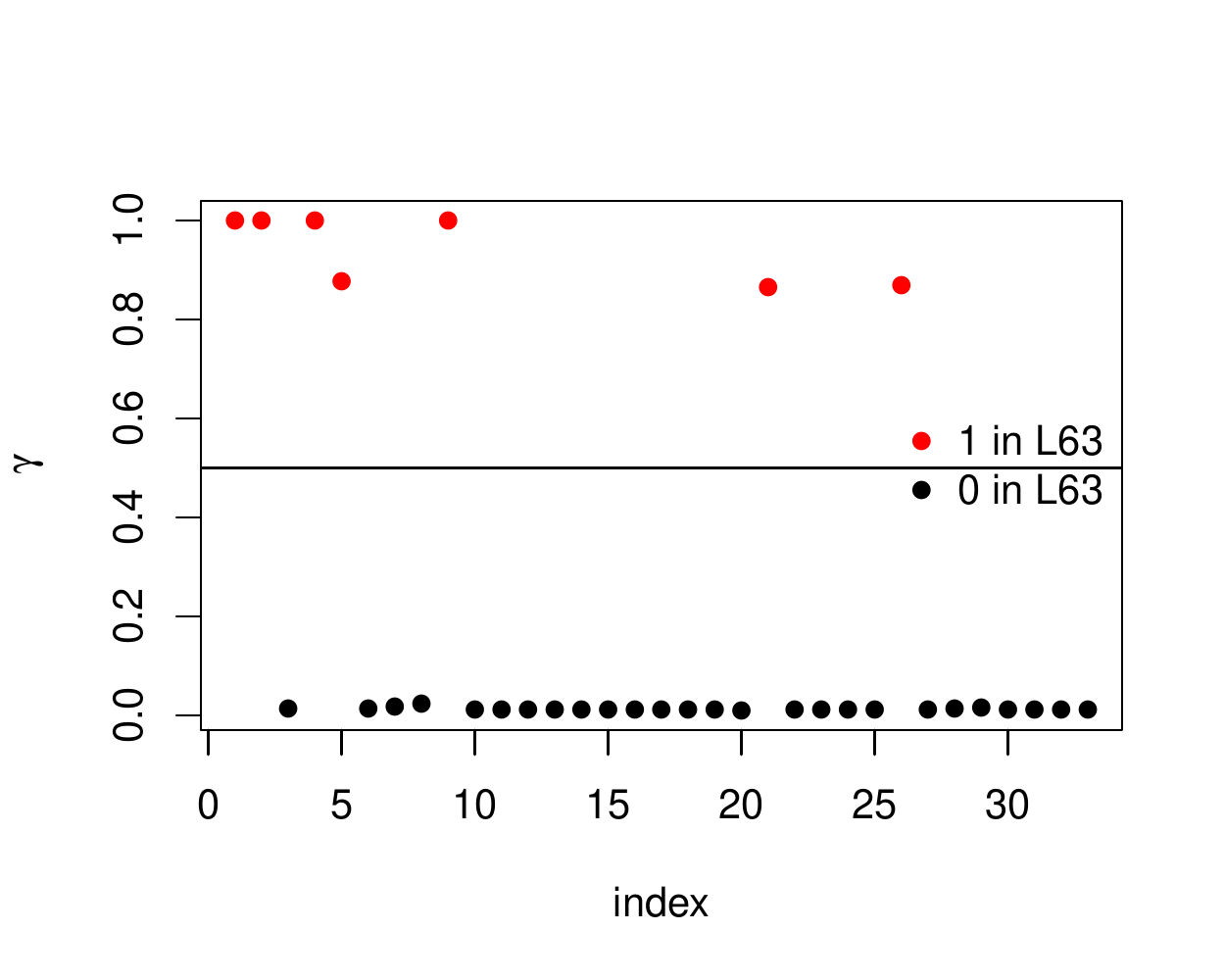}
	\includegraphics[width = 2.8in]{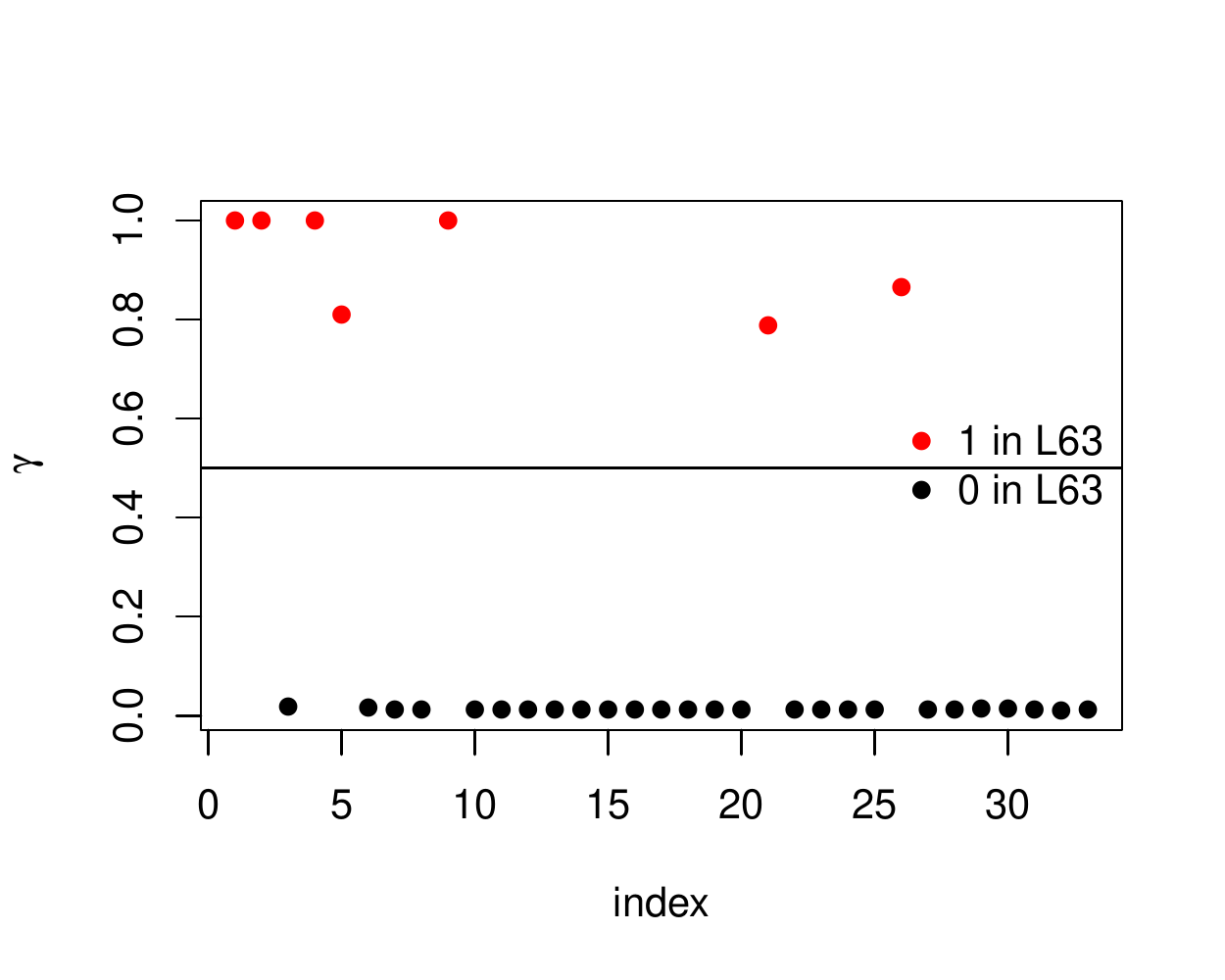}	
	\caption{Posterior mean estimates of $\gamma$ for Lorenz 63, with $\mathbf{X}$ initialized at its truth (left) and with $\mathbf{X}$ initialized at interpolation.}
	\label{fig:l63_gamma}	
\end{figure}

\begin{figure}[htbp]
	\centering
	\includegraphics[width = 2.5in]{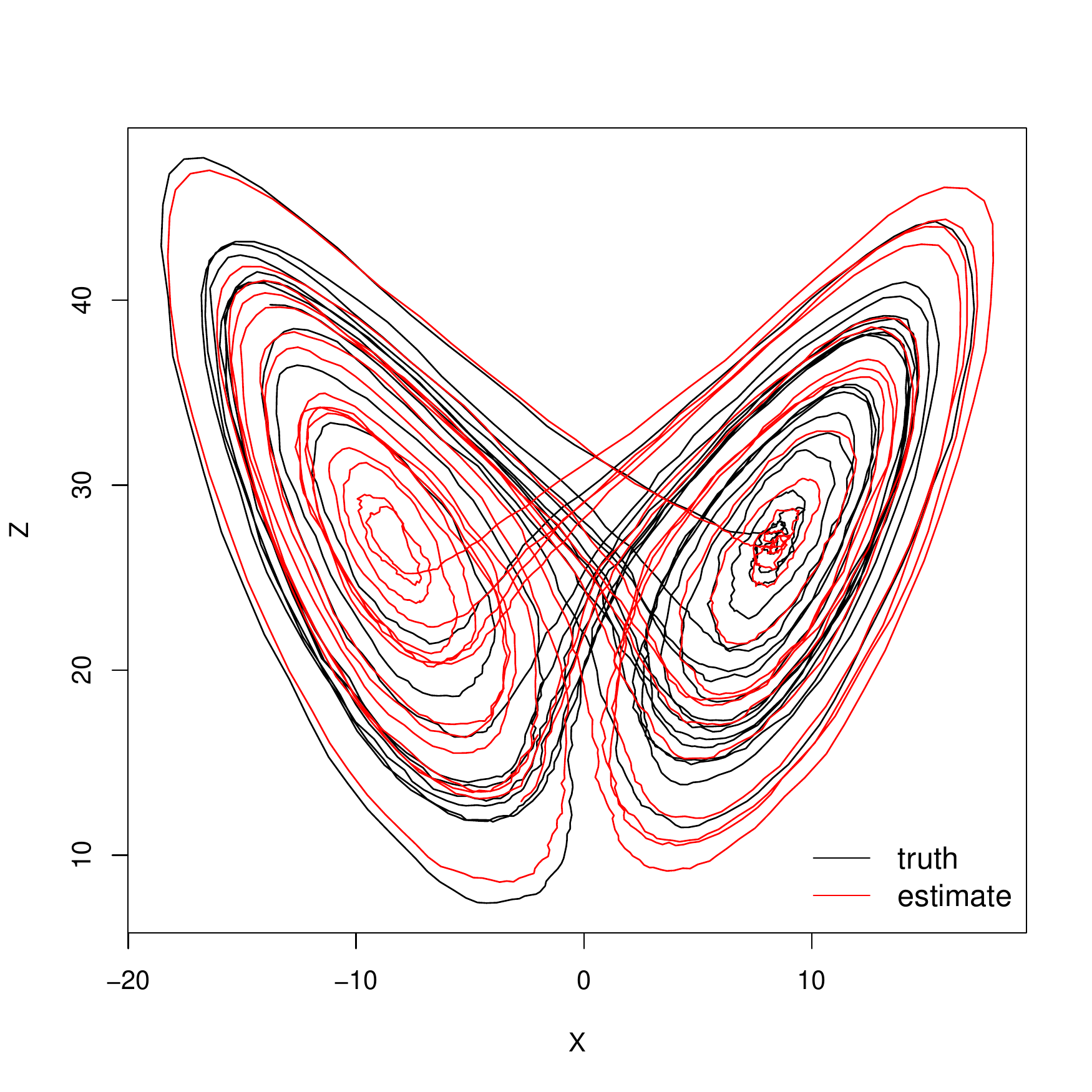}
	\caption{Lorenz-63 trajectory from the truth $(10,28,2.67)$ and estimates $(9.9,27.9,2.66)$.}
	\label{fig:l63_mu_est}	
\end{figure}

The spike-and-slab model for model identification of Lorenz-63 correctly captures the chaotic 
system for samplers initializing \textbf{X} from the truth as well as interpolation with run lengths of $50000$ and 
$10^{5}$ respectively (see Figure~\ref{fig:l63_gamma}). Having identified the system as Lorenz-63, we move on 
the estimation of parameters and latent variables using the linchpin inference model. Similar to the previous 
two examples, we set the prior mean $m_{0}$ to be the parameter estimates from the spike-and-slab model to encapsulate
prior knowledge about the system and run the sampler for $10^{6}$ steps. Note that Lorenz-63 is an extremely 
chaotic system in which slight variations in the drift parameters or $\Sigma$ significantly modify the system trajectory.
The parameter estimates for the sampler with \textbf{X} initialized at interpolations are $(9.9, 27.9, 2.66)$ which
is very close to the true value of $(10,28,2.67)$. Despite this proximity, the L63 trajectory formed by using 
the estimates as the drift parameters differs significantly from the original trajectory (see Figure~\ref{fig:l63_mu_est}).

\begin{figure}[htbp]
	\centering
		\includegraphics[width = 2.2in]{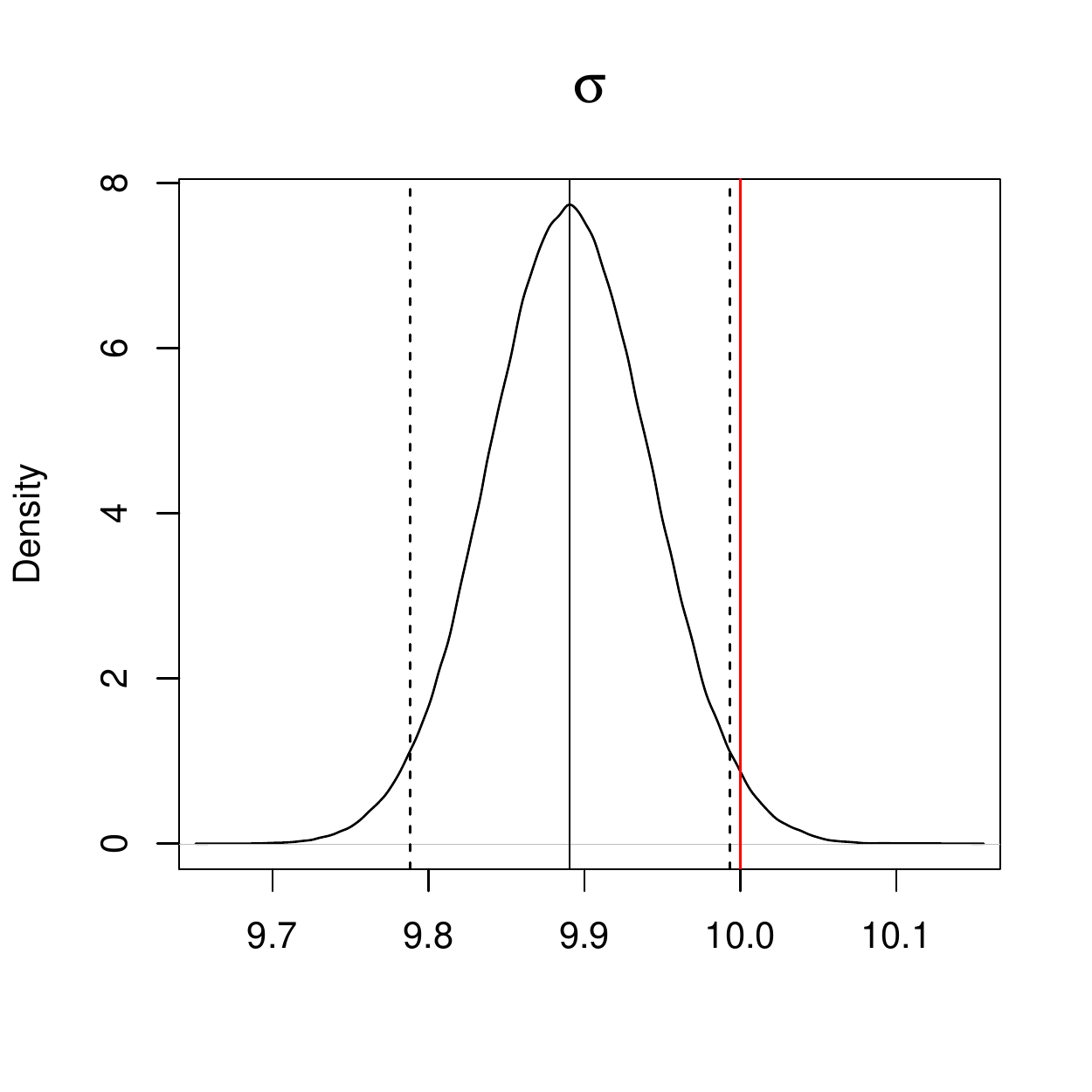}
		\includegraphics[width = 2.2in]{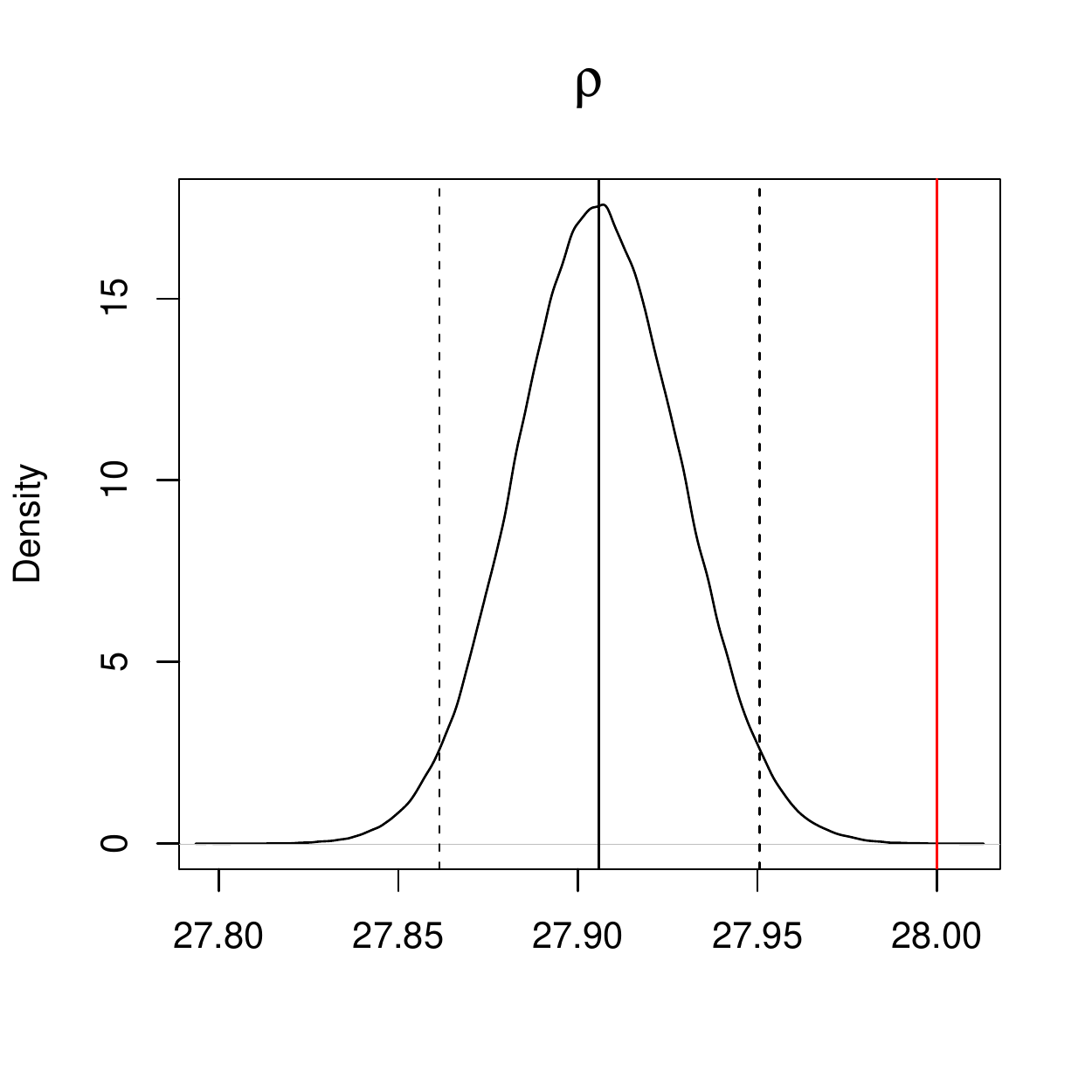}	
		\includegraphics[width = 2.2in]{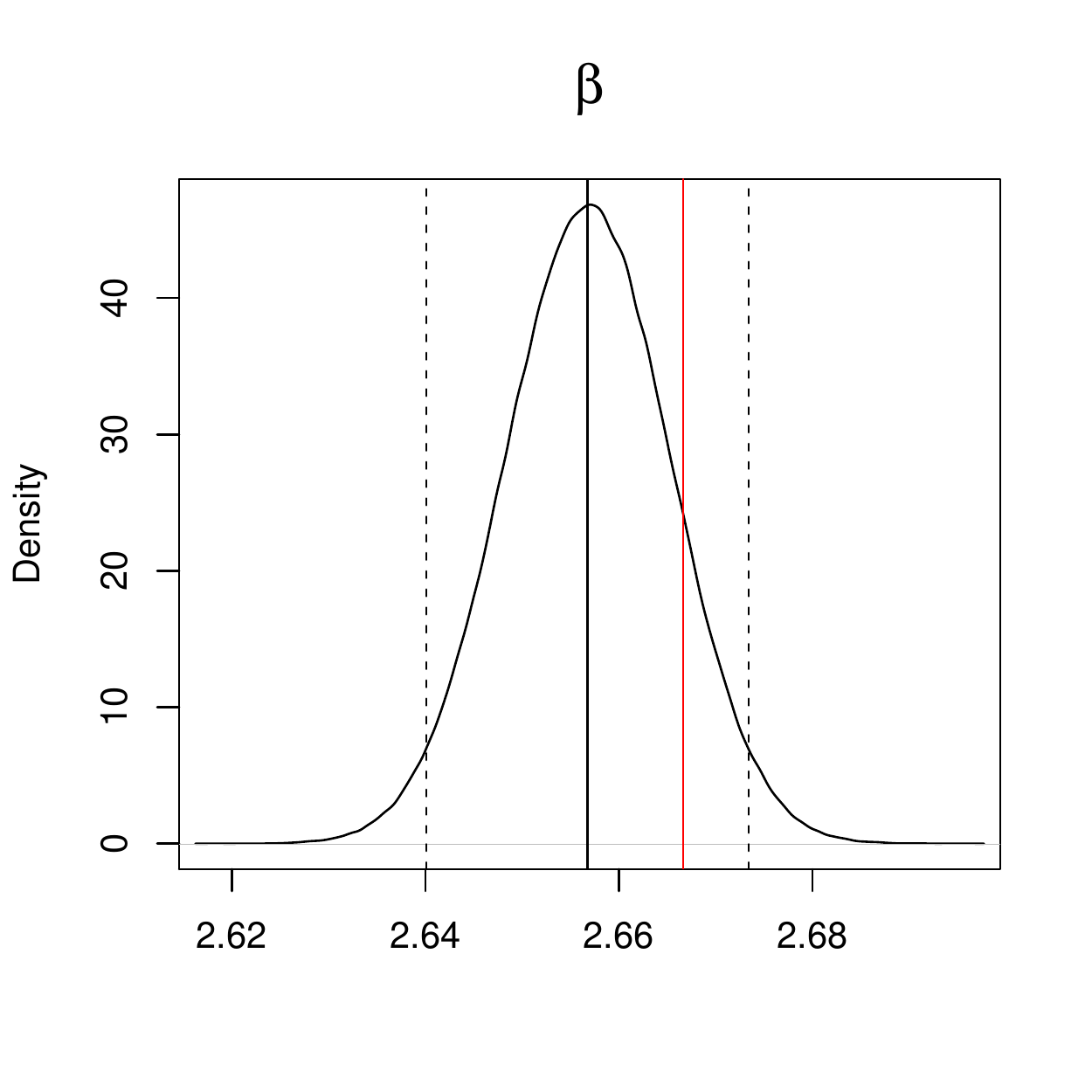}
		\includegraphics[width = 2.2in]{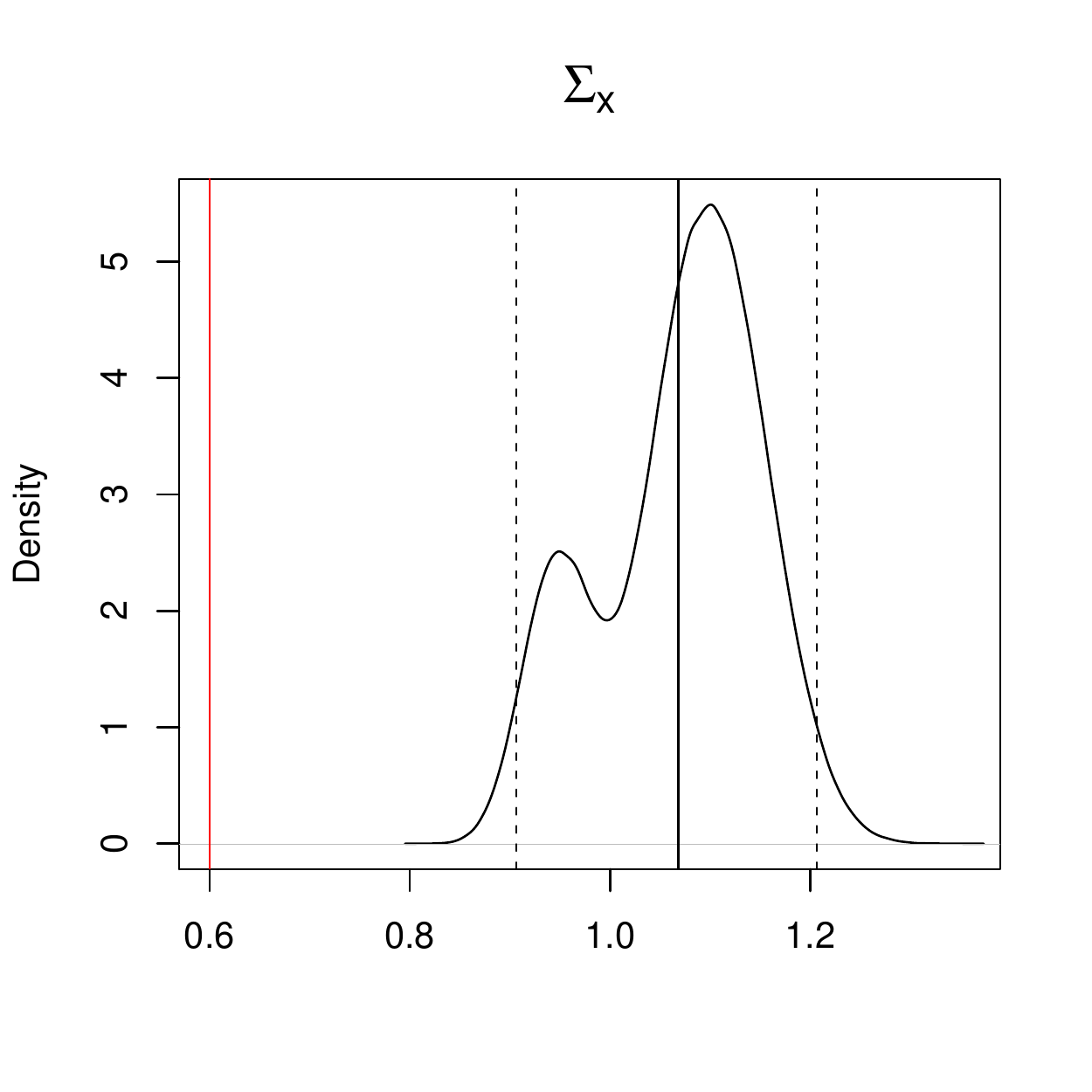}	
		\includegraphics[width = 2.2in]{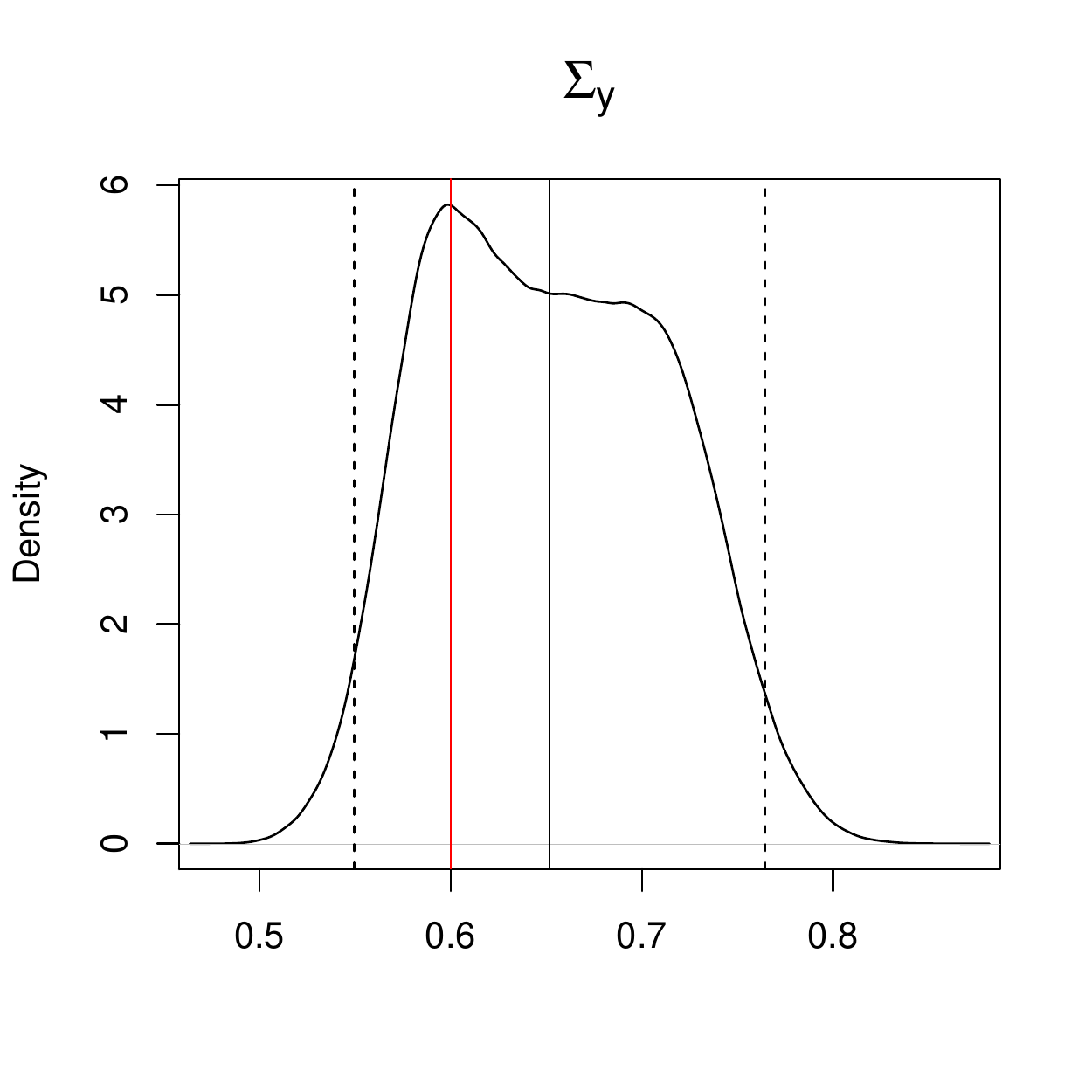}	
		\includegraphics[width = 2.2in]{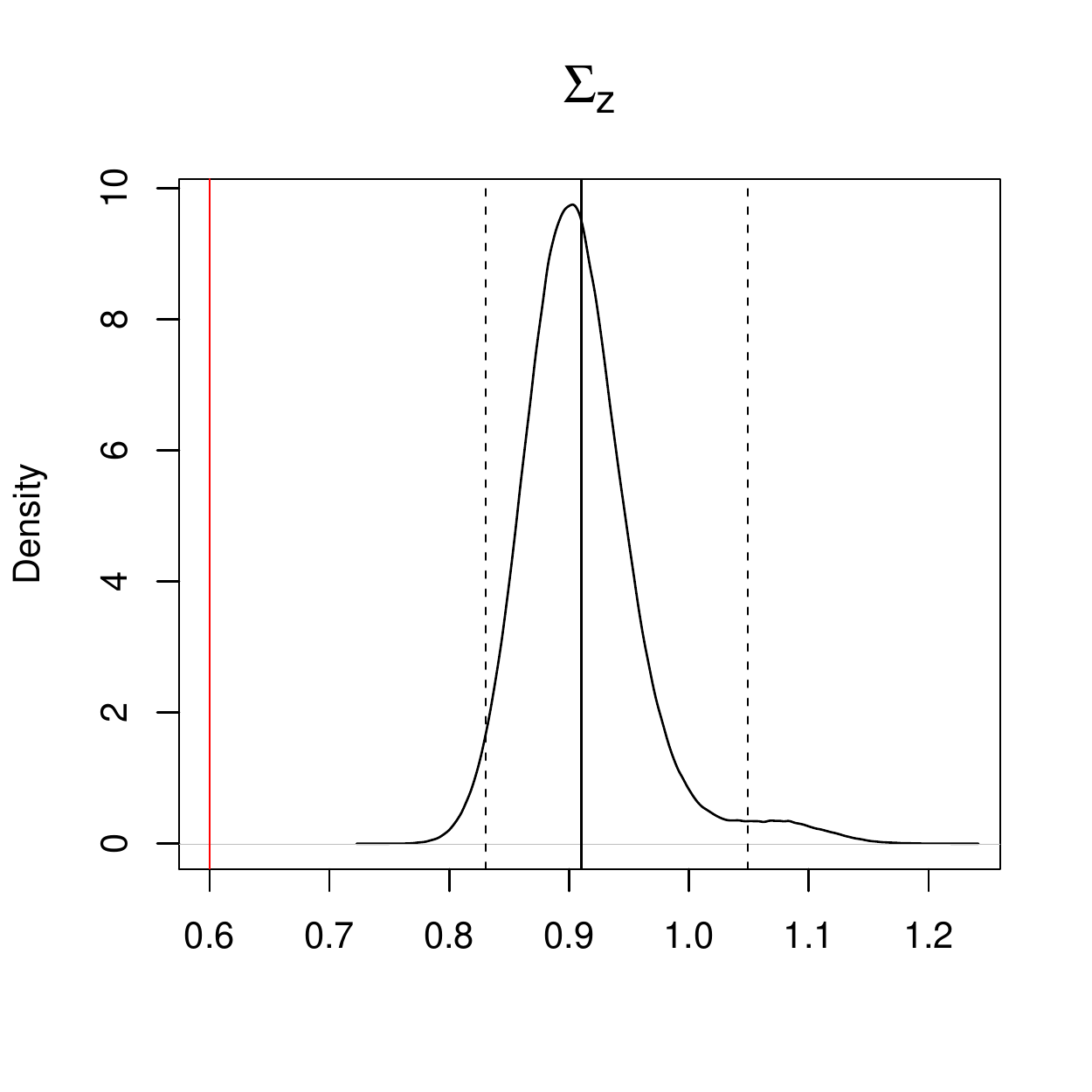}	
		\caption{Lorenz 63 parameter density, starting \textbf{X} at interpolation.}
		\label{fig:l63_interp_inf}			
\end{figure}

\begin{figure}[H]
	\centering
	\begin{subfigure}{0.32\textwidth}
		\centering
		\includegraphics[width = 2in]{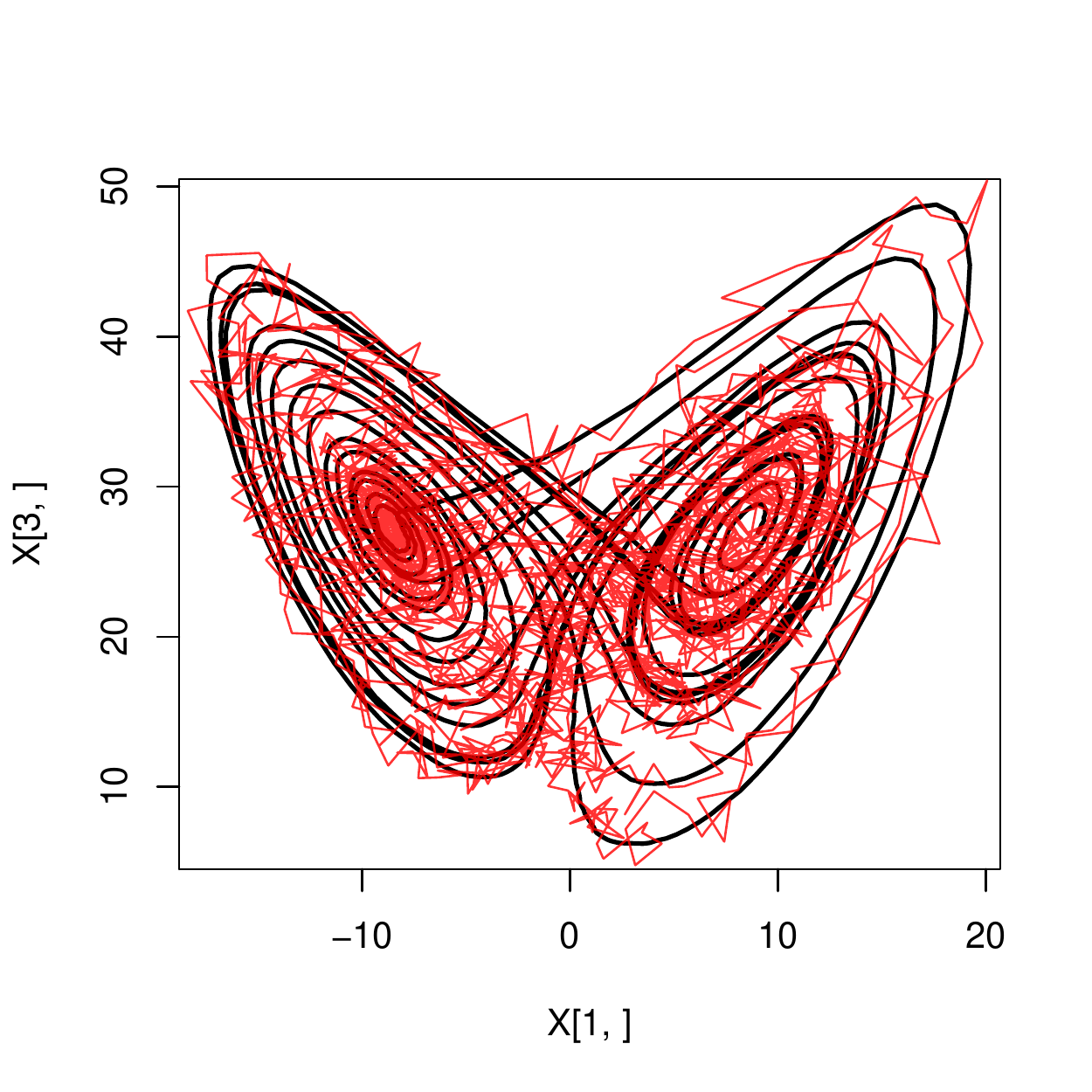}
		\caption{sd = 1}
	\end{subfigure}
	\begin{subfigure}{0.32\textwidth}
		\centering
		\includegraphics[width = 2in]{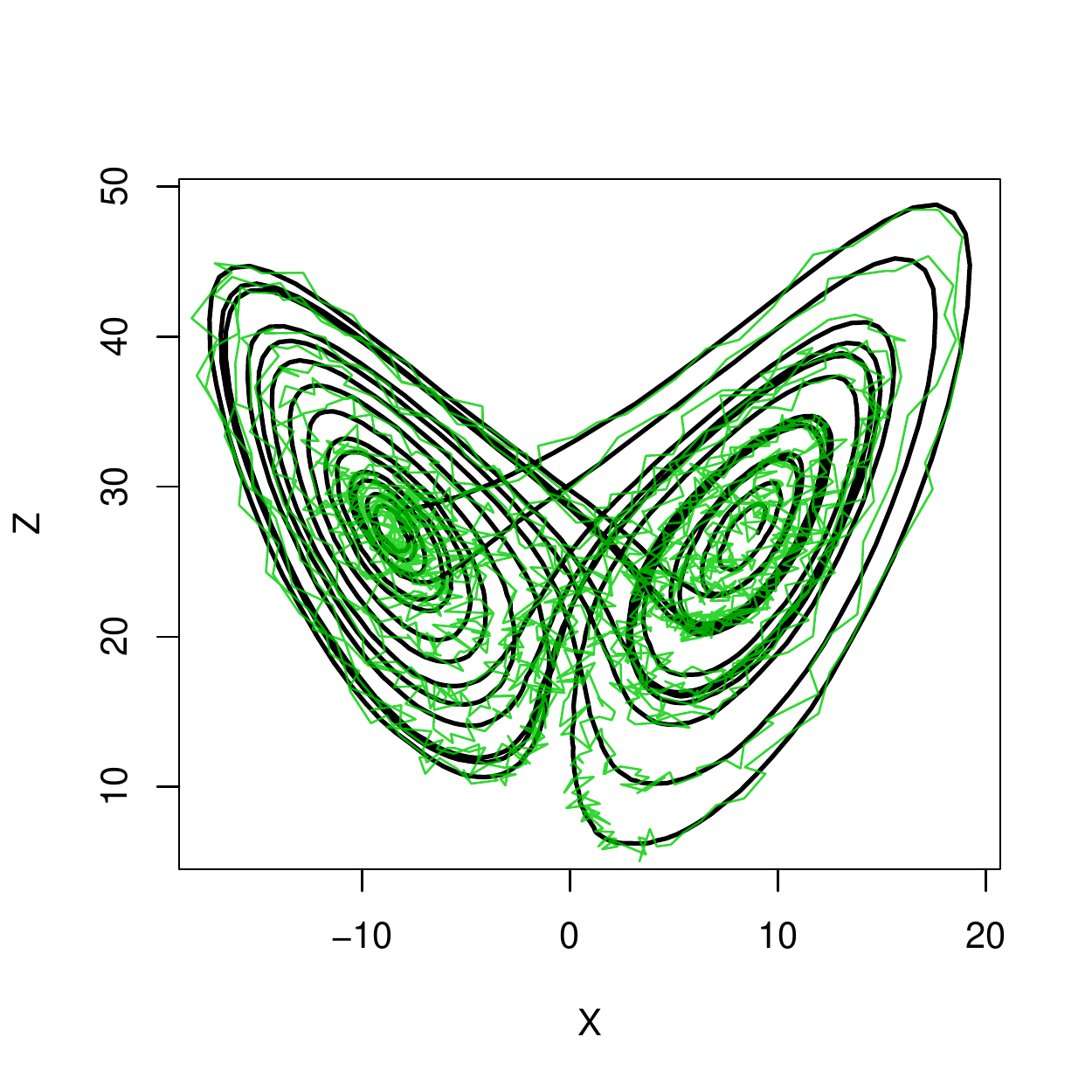}
	\caption{sd = 0.5}
	\end{subfigure}
	\begin{subfigure}{0.32\textwidth}
		\centering
		\includegraphics[width = 2in]{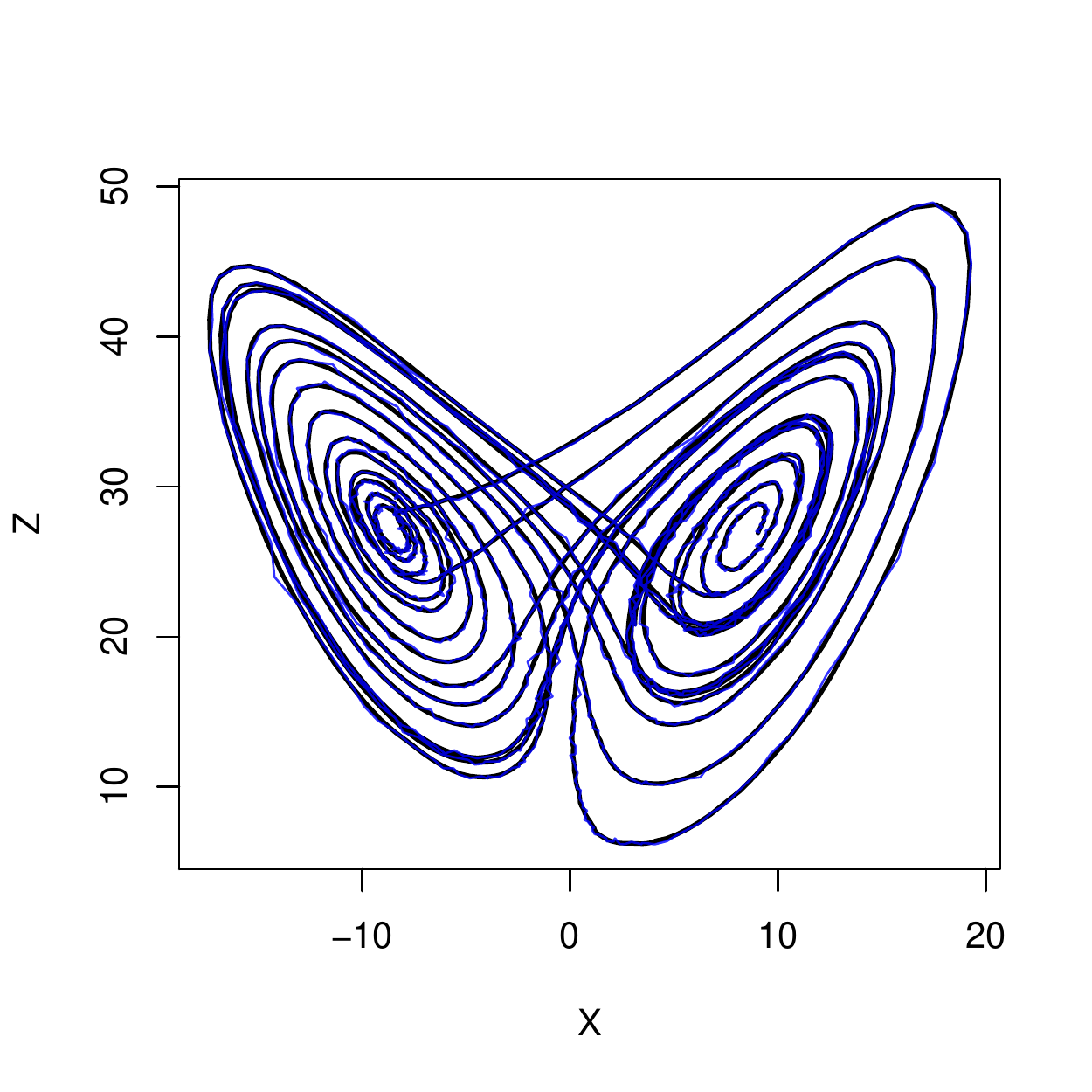}
	\caption{sd = 0.1}
	\end{subfigure}
	\begin{subfigure}{0.48\textwidth}
		\centering
		\includegraphics[width = 2in]{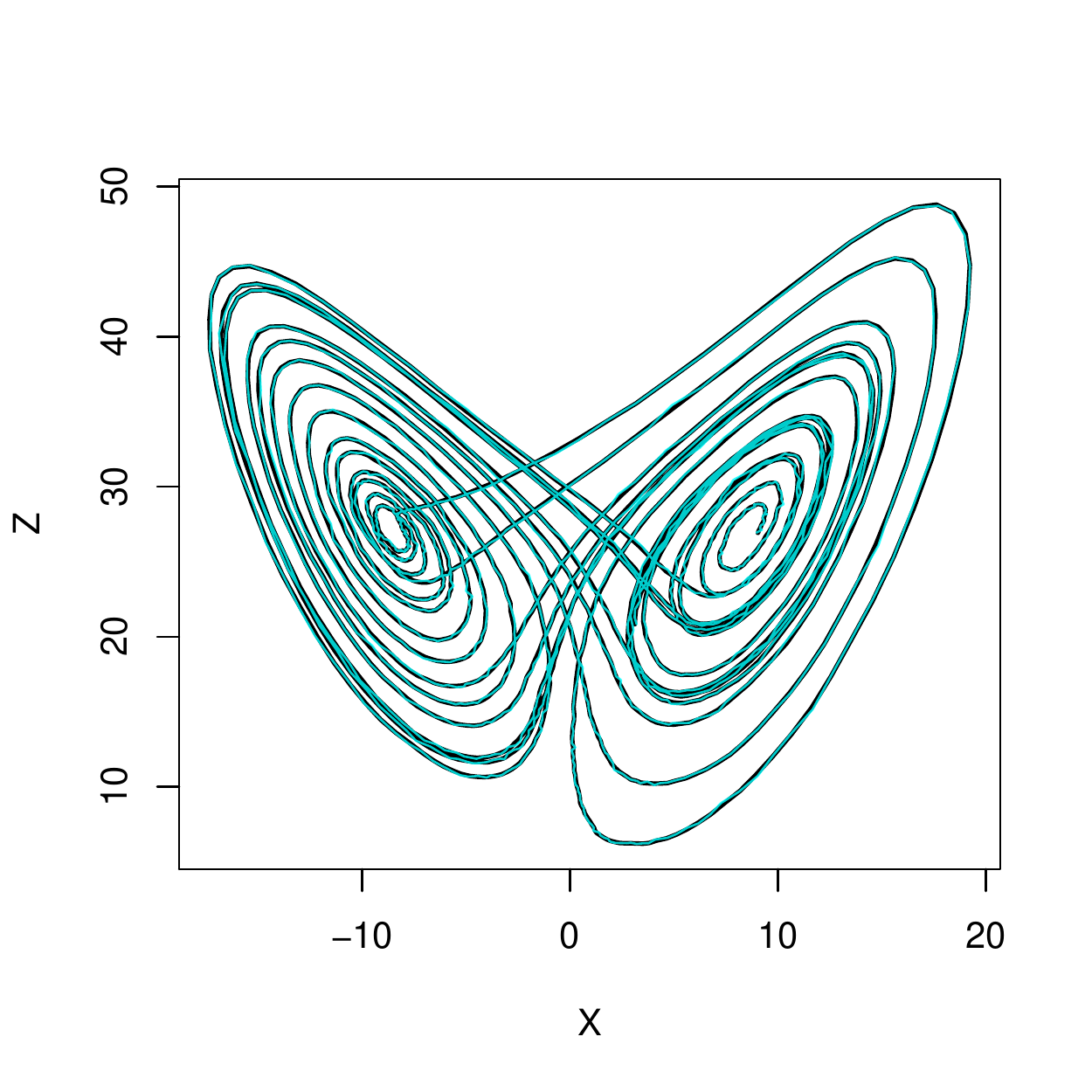}
	\caption{sd = 0.05}
	\end{subfigure}
	\begin{subfigure}{0.48\textwidth}
		\centering
		\includegraphics[width = 2in]{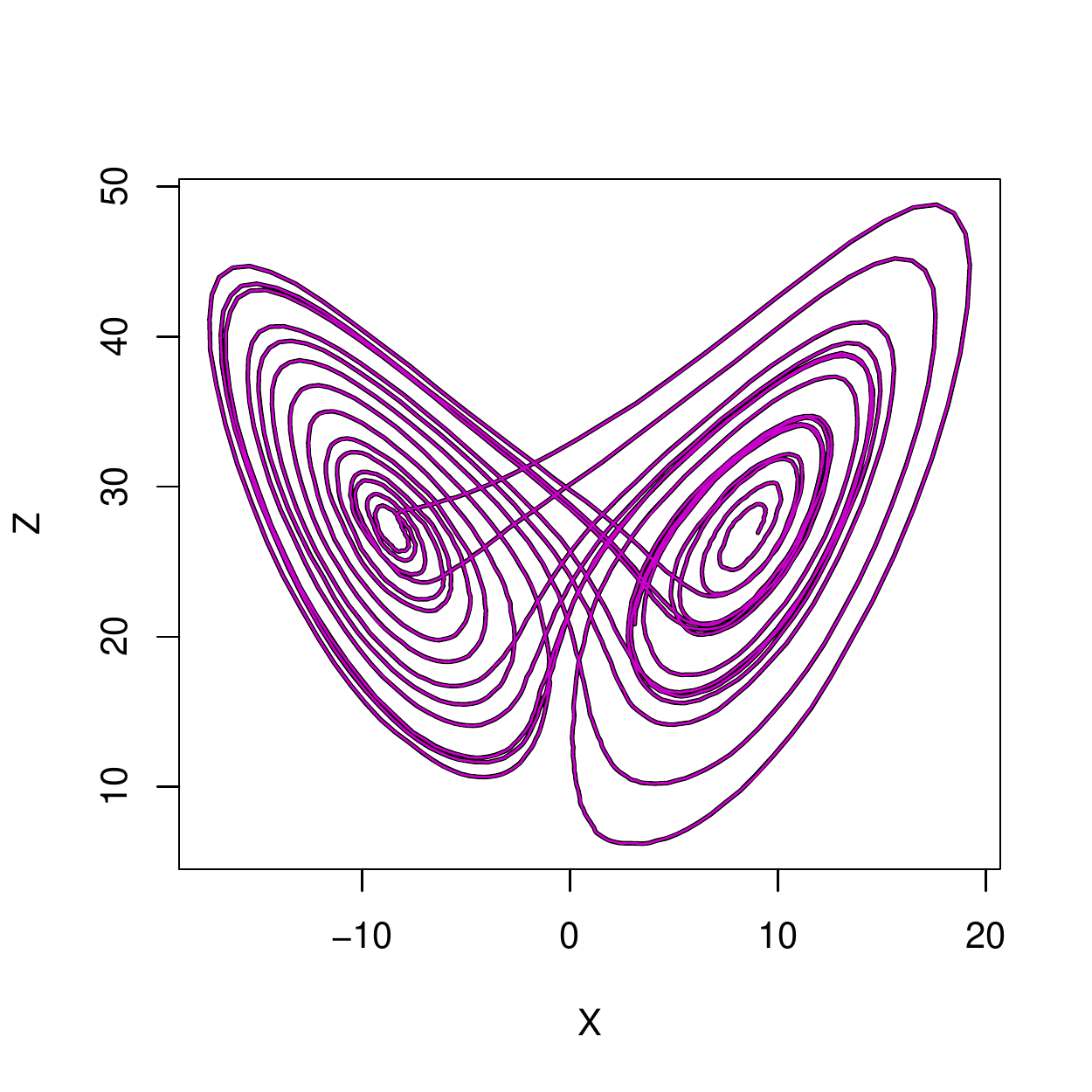}
	\caption{sd = 0.01}
	\end{subfigure}
	\caption{Lorenz-63 trajectory with Gaussian noise}
	\label{fig:butterfly_noise}
\end{figure}

The inference results of $\Sigma$ for the inference run in Figure~\ref{fig:l63_interp_inf} mirror the difficulty in estimating 
\textbf{X} given their chaotic nature and high-dimensionality. To understand the dependence of estimates of $\Sigma$ 
on \textbf{X}, we add Gaussian noise to the Euler-Maruyama approximation of the $\mathbf{X}$ with $\theta = (10,28,8/3)$ and $\Sigma = 0.06$. We use these values to 
calculate the mean of the posterior conditional of $\Sigma$ (Equation~\ref{eq:sigma_full_conditional_inf}). The trajectories 
of \textbf{X} with added noise overlapped 
with the EM approximation in Figure~\ref{fig:butterfly_noise} and corresponding conditional posterior means in Table~\ref{table:butterfly_noise} show that even minor variations in $\mathbf{X}$s offsets $\Sigma$ by orders of magnitude in the Lorenz-63 system. In 
spite of the complications introduced by the relationship between \textbf{X} and $\Sigma$, the drift parameters 
$\theta = [\sigma, \rho, \beta]$ are estimated fairly well owing to the decoupling by the linchpin posterior decomposition.



\section{Future work} 
\label{sec:future_work}

%

In this paper, we assume that the variance of the noise in the observed data, $R$ is known. While this is in keeping with assumptions made in the literature \citep{vrettas:cornford:2011} and a requirement for identifiability, it can be circumvented if additional observations can be taken, for example, if multiple realizations of $Y_{i}$ can be observed at each 
$i = 1, \ldots, K$.

Replacing the continuous time stochastic process $\{ \mathbf{X}_{t} \}$ with the vector $\mathbf{X}$ using the Euler-Maruyama discretization process induces errors in the numeric computations. There are several techniques to address this issue, for example, see \cite{besk:pap:2008,besk:rob:2009, chkrebtii2016bayesian,hennig2015probabilistic, lie2018random, matsuda2019estimation, wang2020role} and references therein. 
The sparsity of data may render some of the alternative techniques from the literature unavailable in the current context.
An alternative may be to use a functional representation of the data, as in \cite{bhaumik2014bayesian, brunel2008parameter,  ramsay2007parameter,zhang2017bayesian} and in several other sources.  We will explore this option in future, but anticipate complications due to the chaotic nature and nonlinear lagged structure of the Lorenz-96 and Lorenz-63 systems.  It has been noted earlier in \cite{vrettas:cornford:2011} that a large number of basis functions are often needed to capture the roughness of the observed physical data. 
Variational algorithms are favored in the related physical sciences literature, and while these have serious shortcomings, they may be useful for obtaining some quantifiers (for example, location parameters) of the posterior distribution relatively quickly, which may help in the actual MCMC computations.

We have adopted a spike-and-slab prior in this paper. Naturally, other Bayesian sparse 
system modeling frameworks, for example using the horseshoe prior, may also be explored 
and studied in this context. The statistical theory related to the methodologies presented here is non-trivial because of the multi-stage computational technique we adopt, and will be developed in a future publication. 

\noindent{\textbf{Acknowledgments:}} (to be filled in for an accepted manuscript).

\bibliographystyle{apalike}
\bibliography{mcref}

\begin{thebibliography}{}

\bibitem[Ala-Luhtala et~al., 2015]{ala2015gaussian}
Ala-Luhtala, J., S{\"a}rkk{\"a}, S., and Pich{\'e}, R. (2015).
\newblock Gaussian filtering and variational approximations for {B}ayesian
  smoothing in continuous-discrete stochastic dynamic systems.
\newblock {\em Signal Processing}, 111:124--136.

\bibitem[Apte et~al., 2007]{apte2007sampling}
Apte, A., Hairer, M., Stuart, A., and Voss, J. (2007).
\newblock Sampling the posterior: An approach to non-{G}aussian data
  assimilation.
\newblock {\em Physica D: Nonlinear Phenomena}, 230(1-2):50--64.

\bibitem[Archambeau et~al., 2007]{archambeau2007variational}
Archambeau, C., Opper, M., Shen, Y., Cornford, D., and Shawe-Taylor, J. (2007).
\newblock Variational inference for diffusion processes.
\newblock {\em Advances in Neural Information Processing Systems}, 20:17--24.

\bibitem[Archila, 2016]{archila:2016}
Archila, F. H.~A. (2016).
\newblock {\em Markov chain {M}onte {C}arlo for Linear Mixed Models}.
\newblock PhD thesis, University of Minnesota.

\bibitem[Batz et~al., 2018]{batz2018approximate}
Batz, P., Ruttor, A., and Opper, M. (2018).
\newblock Approximate {B}ayes learning of stochastic differential equations.
\newblock {\em Physical Review E}, 98:022109.

\bibitem[Bauer et~al., 2017]{bau:gor:2017}
Bauer, S., Gorbach, N.~S., Miladinovic, D., and Buhmann, J.~M. (2017).
\newblock Efficient and flexible inference for stochastic systems.
\newblock In {\em Advances in Neural Information Processing Systems}, pages
  6988--6998.

\bibitem[Beskos et~al., 2009]{besk:rob:2009}
Beskos, A., Papaspiliopoulos, O., and Roberts, G. (2009).
\newblock Monte {C}arlo maximum likelihood estimation for discretely observed
  diffusion processes.
\newblock {\em The Annals of Statistics}, pages 223--245.

\bibitem[Beskos et~al., 2008]{besk:pap:2008}
Beskos, A., Papaspiliopoulos, O., and Roberts, G.~O. (2008).
\newblock A factorisation of diffusion measure and finite sample path
  constructions.
\newblock {\em Methodology and Computing in Applied Probability}, 10:85--104.

\bibitem[Bhaumik and Ghosal, 2014]{bhaumik2014bayesian}
Bhaumik, P. and Ghosal, S. (2014).
\newblock Bayesian estimation in differential equation models.
\newblock {\em arXiv preprint arXiv:1403.0609}.

\bibitem[Brunel et~al., 2008]{brunel2008parameter}
Brunel, N.~J. et~al. (2008).
\newblock Parameter estimation of {ODE}'s via nonparametric estimators.
\newblock {\em Electronic Journal of Statistics}, 2:1242--1267.

\bibitem[Ching et~al., 2006]{ching2006bayesian}
Ching, J., Beck, J.~L., and Porter, K.~A. (2006).
\newblock Bayesian state and parameter estimation of uncertain dynamical
  systems.
\newblock {\em Probabilistic engineering mechanics}, 21:81--96.

\bibitem[Chkrebtii et~al., 2016]{chkrebtii2016bayesian}
Chkrebtii, O.~A., Campbell, D.~A., Calderhead, B., and Girolami, M.~A. (2016).
\newblock Bayesian solution uncertainty quantification for differential
  equations.
\newblock {\em Bayesian Analysis}, 11:1239--1267.

\bibitem[Eraker, 2001]{eraker:2001}
Eraker, B. (2001).
\newblock {MCMC} analysis of diffusion models with application to finance.
\newblock {\em Journal of Business \& Economic Statistics}, 19:177--191.

\bibitem[Flegal et~al., 2020]{mcmcse}
Flegal, J.~M., Hughes, J., Vats, D., and Dai, N. (2020).
\newblock {\em mcmcse: Monte Carlo Standard Errors for {MCMC}}.
\newblock Riverside, CA, Denver, CO, Coventry, UK, and Minneapolis, MN.
\newblock R package version 1.4-1.

\bibitem[George and McCulloch, 1993]{geo:mccul:1993}
George, E.~I. and McCulloch, R.~E. (1993).
\newblock Variable selection via {G}ibbs sampling.
\newblock {\em Journal of the American Statistical Association}, 88:881--889.

\bibitem[Geyer and Johnson, 2020]{geyer:john:2020}
Geyer, C.~J. and Johnson, L.~T. (2020).
\newblock {\em mcmc: Markov Chain {M}onte {C}arlo}.
\newblock R package version 0.9-7.

\bibitem[Hennig et~al., 2015]{hennig2015probabilistic}
Hennig, P., Osborne, M.~A., and Girolami, M. (2015).
\newblock Probabilistic numerics and uncertainty in computations.
\newblock {\em Proceedings of the Royal Society A: Mathematical, Physical and
  Engineering Sciences}, 471(2179):20150142.

\bibitem[Ishwaran et~al., 2005]{ish:2005}
Ishwaran, H., Rao, J.~S., et~al. (2005).
\newblock Spike and slab variable selection: frequentist and {B}ayesian
  strategies.
\newblock {\em The Annals of Statistics}, 33:730--773.

\bibitem[Lie et~al., 2018]{lie2018random}
Lie, H.~C., Sullivan, T.~J., and Teckentrup, A.~L. (2018).
\newblock Random forward models and log-likelihoods in {B}ayesian inverse
  problems.
\newblock {\em SIAM/ASA Journal on Uncertainty Quantification}, 6:1600--1629.

\bibitem[Lorenz, 1963]{lorenz:1963}
Lorenz, E.~N. (1963).
\newblock Deterministic nonperiodic flow.
\newblock {\em Journal of the atmospheric sciences}, 20:130--141.

\bibitem[Lorenz, 1996]{lorenz:1996}
Lorenz, E.~N. (1996).
\newblock Predictability: A problem partly solved.
\newblock In {\em Proc. Seminar on predictability}, volume~1.

\bibitem[Lorenz and Emanuel, 1998]{lorenz:eman:1998}
Lorenz, E.~N. and Emanuel, K.~A. (1998).
\newblock Optimal sites for supplementary weather observations: Simulation with
  a small model.
\newblock {\em Journal of the Atmospheric Sciences}, 55(3):399--414.

\bibitem[Matsuda and Miyatake, 2019]{matsuda2019estimation}
Matsuda, T. and Miyatake, Y. (2019).
\newblock Estimation of ordinary differential equation models with
  discretization error quantification.
\newblock {\em arXiv preprint arXiv:1907.10565}.

\bibitem[Narisetty and He, 2014]{nari:he:2014}
Narisetty, N.~N. and He, X. (2014).
\newblock Bayesian variable selection with shrinking and diffusing priors.
\newblock {\em The Annals of Statistics}, 42(2):789--817.

\bibitem[P{\'e}rez-Vieites et~al., 2018]{perez2018probabilistic}
P{\'e}rez-Vieites, S., Mari{\~n}o, I.~P., and M{\'\i}guez, J. (2018).
\newblock Probabilistic scheme for joint parameter estimation and state
  prediction in complex dynamical systems.
\newblock {\em Physical Review E}, 98:063305.

\bibitem[Ramsay et~al., 2007]{ramsay2007parameter}
Ramsay, J.~O., Hooker, G., Campbell, D., and Cao, J. (2007).
\newblock Parameter estimation for differential equations: a generalized
  smoothing approach.
\newblock {\em Journal of the Royal Statistical Society: Series B (Statistical
  Methodology)}, 69:741--796.

\bibitem[Roberts et~al., 1997]{rob:gel:gilks}
Roberts, G.~O., Gelman, A., Gilks, W.~R., et~al. (1997).
\newblock Weak convergence and optimal scaling of random walk {M}etropolis
  algorithms.
\newblock {\em The Annals of Applied Probability}, 7:110--120.

\bibitem[Ro{\v{c}}kov{\'a} and George, 2018]{rov:george:2018}
Ro{\v{c}}kov{\'a}, V. and George, E.~I. (2018).
\newblock The spike-and-slab lasso.
\newblock {\em Journal of the American Statistical Association}, 113:431--444.

\bibitem[S{\o}rensen, 2004]{soren:2004}
S{\o}rensen, H. (2004).
\newblock Parametric inference for diffusion processes observed at discrete
  points in time: a survey.
\newblock {\em International Statistical Review}, 72:337--354.

\bibitem[Uhlenbeck and Ornstein, 1930]{uhlen:orste:1930}
Uhlenbeck, G.~E. and Ornstein, L.~S. (1930).
\newblock On the theory of the {B}rownian motion.
\newblock {\em Physical review}, 36:823.

\bibitem[Vrettas et~al., 2011]{vrettas:cornford:2011}
Vrettas, M., Cornford, D., and Opper, M. (2011).
\newblock Estimating parameters in stochastic systems: a variational bayesian
  approach.
\newblock {\em Physica D}, 240(23):1877–1900.

\bibitem[Vrettas et~al., 2010]{vrettas2010new}
Vrettas, M.~D., Cornford, D., Opper, M., and Shen, Y. (2010).
\newblock A new variational radial basis function approximation for inference
  in multivariate diffusions.
\newblock {\em Neurocomputing}, 73(7-9):1186--1198.

\bibitem[Wang et~al., 2020]{wang2020role}
Wang, J., Cockayne, J., Oates, C.~J., et~al. (2020).
\newblock A role for symmetry in the {B}ayesian solution of differential
  equations.
\newblock {\em Bayesian Analysis}, 15:1057--1085.

\bibitem[Yu et~al., 2018]{yu:Li:Xu:2018}
Yu, X., Li, J., and Xu, J. (2018).
\newblock Robust adaptive algorithm for nonlinear systems with unknown
  measurement noise and uncertain parameters by variational {B}ayesian
  inference.
\newblock {\em International Journal of Robust and Nonlinear Control},
  28:3475--3500.

\bibitem[Zhang et~al., 2017]{zhang2017bayesian}
Zhang, T., Yin, Q., Caffo, B., Sun, Y., Boatman-Reich, D., et~al. (2017).
\newblock Bayesian inference of high-dimensional, cluster-structured ordinary
  differential equation models with applications to brain connectivity studies.
\newblock {\em The Annals of Applied Statistics}, 11:868--897.

\end{thebibliography}

\begin{table}[H]
  \caption{Mean of posterior conditional of $\Sigma$ for Lorenz 63 with N(0,$s^{2}$) noise added to \textbf{X}. The
  true value of $\Sigma$ for the underlying trajectory is $0.06$.}
  \label{table:butterfly_noise}
  \begin{center}
  \begin{tabular}{|c|c|c|c|c|c|}
    \hline
    Noise variance & $s = 1$ & $s = 0.5$ & $s = 0.1$ & $s = 0.05$ & $s = 0.01$ \\ \hline
    $\Sigma_{x}$ & 185.1 & 44.2 & 2.03 & 0.53 & 0.086 \\ \hline
    $\Sigma_{y}$ & 198.9 & 48.7 & 2.02 & 0.54 & 0.084 \\ \hline
    $\Sigma_{z}$ & 206.8 & 47.3 & 2.04 & 0.52 & 0.084 \\ \hline
  \end{tabular}
  \end{center}
\end{table}

\appendix

\newpage 

\section{Prior and Posterior Calculation}
\label{prior_posterior_calc}

Define $\Theta = (\theta, \Sigma)$ where $\theta \in \mathbb{R}^{d}$ are the parameters of the process and $\Sigma \in \mathbb{R}^{p} \times \mathbb{R}^{p}$ is the diffusion
covariance matrix. Using a $N(\mu_0, \lambda^2_0)$ prior on the first component of \textbf{X} (i.e. $\mathbf{X}_{0}$), the priors for 
the latent variables $\mathbf{X}$ using the Euler-Maruyama approximation can be written as: 
\begin{align*}
    & P(\mathbf{X}_{0:N} \mid \Theta) \\
    &= P(\mathbf{X}_{0}) \prod_{i=1}^{N} P(\mathbf{X}_{i} \mid \mathbf{X}_{i-1}) \\
    &= P(\mathbf{X}_{0}) \prod_{i=1}^{N} \mathbf{N} \left(\mathbf{X}_{i} \mid \mathbf{X}_{i-1} + f(t, \mathbf{X}_{i-1})) \delta t \,, \Sigma \delta t \right) \\
    &= P(\mathbf{X}_{0}) \prod_{i=0}^{N-1} (2 \pi)^{-p/2} |\Sigma \delta t|^{-1/2} \exp\left\{-0.5 \left(\mathbf{X}_{i+1} - \mathbf{X}_{i} - f(t, \mathbf{X}_{i}, \theta) \delta t\right)^{T} \left(\Sigma \delta t\right)^{-1} \right. \\
    & \qquad \qquad  \left(\mathbf{X}_{i+1} - \mathbf{X}_{i} - f(t, \mathbf{X}_{i}, \theta) \delta t\right)\bigg\} \\
    &= P(\mathbf{X}_{0}) (2 \pi)^{-\frac{Np}{2}} |\Sigma \delta t|^{-\frac{N}{2}} \exp\left[\dfrac{- \delta t}{2} \sum_{i=0}^{N-1}\left(\dfrac{\delta \mathbf{X}_{i+1}}{\delta t} - f(t, \mathbf{X}_{i}, \theta)\right)^{T} \Sigma^{-1} \left(\dfrac{\delta \mathbf{X}_{i+1}}{\delta t} - f(t, \mathbf{X}_{i}, \theta)\right)\right] \\
    &= (2 \pi)^{-p/2} |\lambda_{0}^{2}|^{-1/2} \exp\left\{-0.5 (\mathbf{X}_{0} - \mu_{0})^{T} (\lambda_{0}^{2})^{-1}(\mathbf{X}_{0} - \mu_{0})\right\} (2 \pi)^{-Np/2} |\Sigma \delta t|^{-N/2} \\
    & \quad \quad \exp\left[-\dfrac{\delta t}{2} \sum_{i=0}^{N-1}\left(\dfrac{\delta \mathbf{X}_{i+1}}{\delta t} - f(t, \mathbf{X}_{i}, \theta)\right)^{T} \Sigma^{-1} \left(\dfrac{\delta \mathbf{X}_{i+1}}{\delta t} - f(t, \mathbf{X}_{i}, \theta)\right)\right] \,.
\end{align*}
For the $d$ elements of $\theta$, the priors are independent $N(m_{0,j}, s_{0}^{2})$ for $j \in {1,2...,d}$ and the priors for the diagonal elements of 
$\Sigma$ are i.i.d inverse-gamma($\alpha, \beta)$ . \\
The likelihood of the model is given by
\begin{align*}
  P(\mathbf{Y} | \mathbf{X}, \theta, \Sigma) &= \prod_{i=1}^{K} P(\mathbf{Y}_{i} | \mathbf{X}_{t_{i}}, \theta, \Sigma) \\
  &= \prod_{i=1}^{K} (2 \pi)^{-p/2} |R|^{-1/2} \, \exp \left(- \dfrac{(\mathbf{Y}_{i} - \mathbf{X}_{t_{i}})^{T} R^{-1} (\mathbf{Y}_{i} - \mathbf{X}_{t_{i}})}{2}\right) \\
  &= (2 \pi)^{-Kp/2} |R|^{-K/2} \exp \left(-\sum_{i=1}^{K}\dfrac{(\mathbf{Y}_{i} - \mathbf{X}_{t_{i}})^{T} R^{-1} (\mathbf{Y}_{i} - \mathbf{X}_{t_{i}})}{2}\right) \, .
\end{align*}

Therefore, the final form of the posterior can be written as 
\begin{align*}
  & P\left(\mathbf{X}, \Theta | \mathbf{Y}\right) \\ 
  & \propto P(\mathbf{Y} | \mathbf{X}, \theta, \Sigma) \, P(\mathbf{X}) \, P(\Sigma) \, P(\theta) \\
    & \propto (2 \pi)^{-Kp/2} |R|^{-K/2} \exp\left[-0.5 \sum_{i=1}^{K}(\mathbf{Y}_{i} - \mathbf{X}_{t_{i}})^{T}R^{-1}(\mathbf{Y}_{i} - \mathbf{X}_{t_{i}})\right] (2 \pi)^{-p/2} |\lambda_{0}^{2}|^{-1/2} \\
    & \quad \quad  \times \exp\left[-0.5 (\mathbf{X}_{0} - \mu_{0})^{T} (\lambda_{0}^{2})^{-1}(\mathbf{X}_{0} - \mu_{0})\right] (2 \pi)^{-Np/2} |\Sigma \delta t|^{-N/2} \\
    & \quad \quad \times \exp\left[-\dfrac{\delta t}{2} \sum_{i=0}^{N-1}\left(\dfrac{\delta \mathbf{X}_{i+1}}{\delta t} - f(t, \mathbf{X}_{i}, \theta)\right)^{T} \Sigma^{-1} \left(\dfrac{\delta \mathbf{X}_{i+1}}{\delta t} - f(t, \mathbf{X}_{i}, \theta)\right)\right]  \\
    & \quad \quad \times \prod_{i=1}^{d}\left[\exp \left(- \dfrac{(\theta_{i} - m_{0,i})^{2}}{2 s^{2}_{0}}\right)\right] \, \prod_{i=1}^{p}\left[\dfrac{\beta^{\alpha}}{\Gamma(\alpha)}\Sigma_{i}^{-\alpha - 1}\exp\left(-\dfrac{\beta}{\Sigma_{i}}\right)\right]  \\ 
    & \propto \exp \left[-0.5 \left(\sum_{i=1}^{K}(\mathbf{Y}_{i} - \mathbf{X}_{t_{i}})^{T}R^{-1}(\mathbf{Y}_{i} - \mathbf{X}_{t_{i}}) + (\mathbf{X}_{0} - \mu_{0})^{T} (\lambda_{0}^{2})^{-1}(\mathbf{X}_{0} - \mu_{0}) \right. \right. \\
    & \quad \quad + \left. \left. \delta t\sum_{i=0}^{N-1}\left(\dfrac{\delta \mathbf{X}_{i+1}}{\delta t} - f(t, \mathbf{X}_{i}, \theta)\right)^{T} \Sigma^{-1} \left(\dfrac{\delta \mathbf{X}_{i+1}}{\delta t} - f(t, \mathbf{X}_{i}, \theta)\right)\right)\right] \\
    & \quad \quad  \times \prod_{i=1}^{d}\left[\exp \left(- \dfrac{(\theta_{i} - m_{0,i})^{2}}{2 s^{2}_{0}}\right)\right] \, |\Sigma|^{-N/2} \, (\prod_{i=1}^{p}\Sigma_{i})^{-\alpha - 1} \exp \left[-\beta\left(\sum_{i=1}^{p}\dfrac{1}{\Sigma_{i}}\right)\right]\, .
\end{align*}

\section{Spike-and-Slab Posterior Calculation}
\label{spike_slab_calc}

Recall that the prior for  $B$ is,
\begin{align*}
    B_{i,j} | \gamma_{i,j} & \overset{\text{ind}}{\sim} (1 - \gamma_{i,j}) N(0, \tau_{0}^{2}) + \gamma_{i,j} N(0, \tau_{1}^{2}) \, ,\\
    \gamma_{i,j} & \overset{\text{ind}}{\sim} \text{Bernoulli}(q_{i,j}) \, .
\end{align*}

Therefore, the posterior can be written as
\begin{align*}
    & P\left(\mathbf{X}, B, \gamma | \mathbf{Y}\right) \\ 
    & \propto \, \exp \left[-0.5 \left(\sum_{i=1}^{K}(\mathbf{Y}_{i} - \mathbf{X}_{t_{i}})^{T}R^{-1}(\mathbf{Y}_{i} - \mathbf{X}_{t_{i}}) + (\mathbf{X}_{0} - \mu_{0})^{T} (\lambda_{0}^{2})^{-1}(\mathbf{X}_{0} - \mu_{0}) \right. \right. \\
    & \quad \quad + \left. \left. \delta t\sum_{i=0}^{N-1}\left(\dfrac{\delta \mathbf{X}_{i+1}}{\delta t} - B\tilde{\mathbf{X}}_{i}\right)^{T} \Sigma^{-1} \left(\dfrac{\delta \mathbf{X}_{i+1}}{\delta t} - B\tilde{\mathbf{X}}_{i}\right)\right)\right] \, \exp \left[-\beta\left(\dfrac{1}{\Sigma_{x}} +\dfrac{1}{\Sigma_{y}} +\dfrac{1}{\Sigma_{z}}\right)\right]  \\
    & \quad \quad \times (\Sigma_{x} \Sigma_{y} \Sigma_{z})^{-\alpha - 1} \, |\Sigma|^{-N/2}  \, \prod_{i,j}\left[\dfrac{\gamma_{i,j}}{\sqrt{2 \pi \tau_{1}^{2}}} \, \exp \left(-\dfrac{B_{i,j}^{2}}{2 \tau_{1}^{2}}\right) + \dfrac{1 - \gamma_{i,j}}{\sqrt{2 \pi \tau_{0}^{2}}} \, \exp \left(- \dfrac{B_{i,j}^{2}}{2 \tau_{0}^{2}}\right) \right] \\
    & \quad \quad \times \prod_{i,j} (q_{i,j})^{\gamma_{i,j}} (1 - q_{i,j})^{1 - \gamma_{i,j}} \, .
\end{align*}
We can construct a linchpin sampler for this posterior with $(\mathbf{X}, B, \gamma)$ as the linchpin variable. For 
$i = [1,2,3]$
\begin{align*}
    \Sigma_{i} | \mathbf{X}, \theta, \mathbf{Y} \sim \text{inv-gamma} \left(\dfrac{N}{2} + \alpha, \, \beta + \dfrac{\delta t}{2}\, \sum_{j=0}^{N-1} \left(\dfrac{\delta \mathbf{X}_{j+1}}{\delta t} - B\tilde{\mathbf{X}}_{j}\right)^{2}_{i}\right) \, .
\end{align*}

Define $Q(B, \mathbf{X})$ as 
\begin{align*}
    Q(B, \mathbf{X}) &=  C_{5} \exp \left[-0.5 \left(\sum_{i=1}^{K}(\mathbf{Y}_{i} - \mathbf{X}_{t_{i}})^{T}R^{-1}(\mathbf{Y}_{i} - \mathbf{X}_{t_{i}}) + (\mathbf{X}_{0} - \mu_{0})^{T} (\lambda_{0}^{2})^{-1}(\mathbf{X}_{0} - \mu_{0}) \right)\right] \\
    & \quad \quad \times \prod_{i,j}\left[\dfrac{\gamma_{i,j}}{\sqrt{2 \pi \tau_{1}^{2}}} \, \exp \left(-\dfrac{B_{i,j}^{2}}{2 \tau_{1}^{2}}\right) + \dfrac{1 - \gamma_{i,j}}{\sqrt{2 \pi \tau_{0}^{2}}} \, \exp \left(- \dfrac{B_{i,j}^{2}}{2 \tau_{0}^{2}}\right) \right] \, .
\end{align*}

Following the steps from the next section, we can write the marginal of the linchpin variable as
\begin{align*}
    P(\mathbf{X}, B, \gamma | \mathbf{Y}) &= \int_{\mathbb{R}_{+}^{3}} P(\mathbf{X}, B, \Sigma | \mathbf{Y}) \, d \Sigma \\
    &= \dfrac{Q(B, \mathbf{X}) \, \left(\Gamma(\alpha + N/2)\right)^{3}\, \prod_{i,j} (q_{i,j})^{\gamma_{i,j}} (1 - q_{i,j})^{1 - \gamma_{i,j}}}{\prod_{p=1}^{3}\left(\beta + \dfrac{\delta t}{2}\, \sum_{i=0}^{N-1} \left(\dfrac{\delta \mathbf{X}_{i+1}}{\delta t} - B \tilde{\mathbf{X}}_{i}\right)^{2}_{p}\right)^{\alpha+N/2}} \, .
\end{align*}

We can construct a Metropolis-within-Gibbs sampler to sample from $P(\mathbf{X}, B, \gamma | \mathbf{Y})$, 
separating $\gamma_{i,j}$ and $(\mathbf{X}, B)$ into their conditionals. Following this scheme, we will first 
sample the discrete random variable $\gamma$ using its conditional and use MH to sample $(\mathbf{X}, B)$ from 
its conditional with the updated values of $\gamma$.
\begin{align*}
    P(\gamma_{i,j} = 1 | \mathbf{X}, B, \gamma_{(i,j)}, \mathbf{Y}) & \propto \left(\dfrac{1}{\sqrt{2 \pi \tau_{1}^{2}}} \exp \left(- \dfrac{B_{i,j}^{2}}{2 \tau_{1}^{2}}\right)\right)(q_{i,j}) \,. \\
    P(\gamma_{i,j} = 0 | \mathbf{X}, B, \gamma_{(i,j)}, \mathbf{Y}) & \propto \left(\dfrac{1}{\sqrt{2 \pi \tau_{0}^{2}}} \exp \left(- \dfrac{B_{i,j}^{2}}{2 \tau_{0}^{2}}\right)\right)(1-q_{i,j}) \, .
\end{align*}

Here, $\gamma_{(i,j)}$ denotes all elements of $\gamma$ except the $(i,j)th$ element. The conditional for the remaining parameters is
\begin{align*}
    P(\mathbf{X}, B | , \gamma, \mathbf{Y}) &= \dfrac{Q(B, \mathbf{X}) \, \left(\Gamma(\alpha + N/2)\right)^{3}}{\prod_{p=1}^{3}\left(\beta + \dfrac{\delta t}{2}\, \sum_{i=0}^{N-1} \left(\dfrac{\delta \mathbf{X}_{i+1}}{\delta t} - B \tilde{\mathbf{X}}_{i}\right)^{2}_{p}\right)^{\alpha+N/2}} \, .
\end{align*}

Since this doesn't follow any known distribution, we will use MH to generate samples from this conditional. 

\section{Linchpin Sampler}
\label{sec:linchpin_theory}
The posterior is given by
\[
P(\mathbf{X}, \theta, \Sigma | \mathbf{Y}) = P(\Sigma | \mathbf{Y}, \theta, \mathbf{X}) \, P(\mathbf{X}, \theta | \mathbf{Y}) \, .
\]
A linchpin sampler decomposes the posterior into the product of a conditional and a marginal (called the linchpin variable).
Here, we treat $(\mathbf{X}, \theta)$ as the linchpin variable.

Define $Q(\theta, \mathbf{X})$ as 
\begin{align*}
    Q(\theta, \mathbf{X}) &=  C_{5} \exp \left[-0.5 \left(\sum_{i=1}^{K}(\mathbf{Y}_{i} - \mathbf{X}_{t_{i}})^{T}R^{-1}(\mathbf{Y}_{i} - \mathbf{X}_{t_{i}}) + (\mathbf{X}_{0} - \mu_{0})^{T} (\lambda_{0}^{2})^{-1}(\mathbf{X}_{0} - \mu_{0}) \right)\right] \\
    & \quad \quad \times \prod_{i=1}^{d}\left[\exp \left(- \dfrac{(\theta_{i} - m_{0,i})^{2}}{2 s^{2}_{0}} \right)\right]  \, .
\end{align*}
We can integrate out $\Sigma$ out of the posterior to obtain the marginal of the linchpin variable. 
\begin{align*}
  & P(\mathbf{X}, \theta | \mathbf{Y}) \\
  &= \int_{\mathbb{R}_{+}^{p}} P(\mathbf{X}, \theta, \Sigma | \mathbf{Y}) \, d \Sigma \\
    &= \int_{\mathbb{R}_{+}^{p}} C_{5} \exp \left[-0.5 \left(\sum_{i=1}^{K}(\mathbf{Y}_{i} - \mathbf{X}_{t_{i}})^{T}R^{-1}(\mathbf{Y}_{i} - \mathbf{X}_{t_{i}}) + (\mathbf{X}_{0} - \mu_{0})^{T} (\lambda_{0}^{2})^{-1}(\mathbf{X}_{0} - \mu_{0}) \right. \right. \\
    & \quad \quad + \left. \left. \delta t\sum_{i=0}^{N-1}\left(\dfrac{\delta \mathbf{X}_{i+1}}{\delta t} - f(t, \mathbf{X}_{i}, \theta)\right)^{T} \Sigma^{-1} \left(\dfrac{\delta \mathbf{X}_{i+1}}{\delta t} - f(t, \mathbf{X}_{i}, \theta)\right)\right)\right] \, \prod_{i=1}^{d}\left[\exp \left(- \dfrac{(\theta_{i} - m_{0,i})^{2}}{2 s^{2}_{0}} \right)\right] \\
    & \quad \quad  |\Sigma|^{-N/2} \, (\prod_{i=1}^{p}\Sigma_{i})^{-\alpha - 1} \exp \left[-\beta\left(\sum_{i=1}^{p}\dfrac{1}{\Sigma_{i}}\right)\right] \, d \Sigma \\
    &= \int_{\mathbb{R}_{+}^{p}} Q(\theta, \mathbf{X}) \, |\Sigma|^{-N/2} (\prod_{i=1}^{p}\Sigma_{i})^{-\alpha-1} \, \exp \left[-\beta\left(\sum_{i=1}^{p}\dfrac{1}{\Sigma_{i}}\right)\right]  \\
    & \quad \quad \exp\left[- \dfrac{\delta t}{2}\left(\sum_{i=0}^{N-1}\left(\dfrac{\delta \mathbf{X}_{i+1}}{\delta t} - f(t, \mathbf{X}_{i}, \theta)\right)^{T} \Sigma^{-1} \left(\dfrac{\delta \mathbf{X}_{i+1}}{\delta t} - f(t, \mathbf{X}_{i}, \theta)\right)\right)\right] \, d \Sigma \\
    &= \int_{\mathbb{R}_{+}^{p}} Q(\theta, \mathbf{X}) \prod_{i=1}^{p} \left[ \Sigma_{i}^{-\alpha - N/2 - 1} \, \exp\left(-\left[\beta + \dfrac{\delta t}{2}\, \sum_{j=0}^{N-1} \left(\dfrac{\delta \mathbf{X}_{j+1}}{\delta t} - f(t, \mathbf{X}_{j}, \theta\right)^{2}_{i}\right] / \Sigma_{i}\right) \right]\, d \Sigma \\
    &=  \dfrac{Q(\theta, \mathbf{X}) \, \left(\Gamma(\alpha + N/2)\right)^{p}}{\prod_{i=1}^{p}\left(\beta + \dfrac{\delta t}{2}\, \sum_{j=0}^{N-1} \left(\dfrac{\delta \mathbf{X}_{j+1}}{\delta t} - f(t, \mathbf{X}_{j}, \theta\right)^{2}_{i}\right)} \int_{\mathbb{R}_{+}^{p}} \prod_{i=1}^{p} \text{inv-gamma}(\alpha_{i}, \beta{i}) \, d \Sigma \\
    &= \dfrac{Q(\theta, \mathbf{X}) \, \left(\Gamma(\alpha + N/2)\right)^{p}}{\prod_{i=1}^{p}\left(\beta + \dfrac{\delta t}{2}\, \sum_{j=0}^{N-1} \left(\dfrac{\delta \mathbf{X}_{j+1}}{\delta t} - f(t, \mathbf{X}_{j}, \theta)\right)^{2}_{i}\right)^{\alpha+N/2}} \, .
\end{align*}

Using this decomposition, we can generate a sample from the linchpin variable and use that to get a sample from $\Sigma$ using 
\begin{equation}
  \label{eqn:Sigma_cond_post}
  \Sigma_{i} | \mathbf{X}, \theta, \mathbf{Y} \sim \text{inv-gamma} \left(\dfrac{N}{2} + \alpha, \, \beta + \dfrac{\delta t}{2}\, \sum_{j=0}^{N-1} \left(\dfrac{\delta \mathbf{X}_{j+1}}{\delta t} - f(t, \mathbf{X}_{j}, \theta) \right)^{2}_{i}\right) \, ,
\end{equation}
for $i \in {1,2,...,p}$.

\end{document}